\documentclass[aps,pre,twocolumn,showpacs,groupedaddress]{revtex4}
\usepackage{latexsym}
\usepackage{graphicx}
\usepackage{amsmath}
\usepackage{amssymb}
\usepackage{bm}
\begin{document}
\title{Microscopic theory of glassy dynamics and 
  glass transition for molecular crystals}
\author{Michael Ricker}
\email{mricker@uni-mainz.de}
\author{Rolf Schilling}
\email{rschill@uni-mainz.de}
\affiliation{Institut f\"ur Physik, Johannes Gutenberg-Universit\"at
Mainz, Staudinger Weg 7, D-55099 Mainz, Germany}
\date{\today}
\begin{abstract}
We derive a microscopic equation of motion for the dynamical
orientational correlators of molecular crystals. Our approach is
based upon mode coupling theory. Compared to liquids we find four main 
differences: (i) the memory kernel contains Umklapp processes if the 
total momentum of two orientational modes is outside the first Brillouin 
zone, (ii) besides the static two-molecule orientational correlators 
one also needs the static one-molecule orientational density as an 
input, where the latter is nontrivial due to the crystal's anisotropy, 
(iii) the static orientational current density correlator does 
contribute an anisotropic, inertia-independent part to the memory 
kernel, (iv) if the molecules are assumed to be fixed on a rigid 
lattice, the tensorial orientational correlators and the memory 
kernel have vanishing $l,l'=0$ components, due to the absence 
of translational motion. The resulting mode coupling equations are 
solved for hard ellipsoids of revolution on a rigid sc-lattice. 
Using the static orientational correlators from Percus-Yevick 
theory we find an ideal glass transition generated due to
precursors of orientational order which depend on $X_{0}$ and 
$\varphi$, the aspect ratio and packing fraction of the ellipsoids.
The glass formation of oblate ellipsoids is enhanced compared 
to that for prolate ones. For oblate ellipsoids with $X_{0} \lesssim 
0.7$ and prolate ellipsoids with $X_{0} \gtrsim 4$, the 
critical diagonal nonergodicity parameters in reciprocal space exhibit 
more or less sharp maxima at the zone center with very small values 
elsewhere, while for prolate ellipsoids with $2 \lesssim X_{0} 
\lesssim 2.5$ we have maxima at the zone edge. The off-diagonal 
nonergodicity parameters are not restricted to positive values and 
show similar behavior. For $0.7 \lesssim X_{0} \lesssim 2$, no 
glass transition is found because of too small static orientational 
correlators. In the glass phase, the nonergodicity parameters show 
a much more pronounced ${\bf q}$-dependence.
\end{abstract}
\maketitle
\section{Introduction}
\label{secI}
The experimental and theoretical investigation of systems with
\textit{self-generated} disorder has mainly been devoted to simple
and molecular liquids \cite{Proc1,Spe03}. In their supercooled
state at low temperatures or high densities, liquids exhibit
nontrivial dynamics, often called glassy dynamics. Decreasing the
temperature $T$ or increasing the number density $n$ may result in
a glass transition. The physical origin of glassy dynamics and the
glass transition is the formation of a cage by the particles. For
not too low temperatures and not too high densities, the cage's
lifetime is finite, i.e.~a particle can escape with a finite
probability. However, if the lifetime diverges, e.g. at a critical
temperature, the particles remain localized in their cages. In
that case an ideal glass transition occurs at that temperature. 

There is no general agreement about the theoretical
description of the glass transition. From a practical point of
view one often considers the so-called calorimetric glass
transition temperature $T_g$ as the temperature at which a
supercooled liquid becomes a structural glass. But $T_g$ depends
on the cooling process. Therefore, it is not well defined.
Besides $T_g$, there exist two better defined characteristic
temperatures, which are $T_c$, the mode coupling transition
temperature, and $T_K$, the Kauzmann temperature. At $T_K$ the
excess entropy of a supercooled liquid with respect to its
crystalline phase disappears. This is a very old concept which
only recently has been put onto a microscopic basis by the replica
theory for structural glasses (see \cite{Mezo02} and references
therein). The existence of a purely dynamical glass transition at
a critical temperature
$T_c$ has been suggested about two decades ago \cite{UBen84}. This
approach is based on mode coupling theory (MCT). MCT describes the
cage effect (as explained above) in a self-consistent way. At
$T_c$ a transition from an ergodic to a nonergodic phase occurs.
Close to $T_c$ the relaxational dynamics of e.g. the density
fluctuations, exhibits two time scaling laws with relaxation times
which diverge at $T_c$. For more details and comparison of the MCT
predictions with experimental and simulational results the reader
is referred to Refs. \cite{WGoe91,RSRi54,WGoe98,WGoe99,KBin03}. A
review of MCT, replica theory of structural glasses and a
selection of phenomenological theories is given in Ref.
\cite{RS04}.

Glassy behavior of systems with self-generated disorder is not
restricted to liquids. There exist so-called molecular crystals
\cite{JDW01} were molecules are located at sites of a periodic 
lattice. At higher temperatures their orientational degrees of
freedom (ODOF) may be dynamically disordered, i.e.~ergodic. This
phase is called the plastically crystalline phase \cite{Past79}.
Decreasing temperature lowers the lattice constants which in turn
leads to an increase of the steric hindrance between the ODOF.
This may result in the formation of an ``orientational'' cage in
which the orientation of a molecule is captured on a certain time
scale, quite similar to liquids. If this time scale diverges the
plastic crystal undergoes an ideal orientational glass transition.
The corresponding phase is called \textit{glassy crystal}. That such
a glass transition really occurs has been proven experimentally
several decades ago. First systematic experimental indication for the
formation of glassy crystals has been given in 1974 for several
molecular crystals \cite{HSU74}. Since then a lot of glassy
crystals were found. Without claiming completeness the most
intensively studied molecular crystals are cyanoadamantane
\cite{JLSan82,RBranLu,AffDes99,JKoga00}, chloradamantane
\cite{AffCoDe01} and ethanol \cite{ASriBe96,RamVie97}. Ethanol has
the big advantage that it can form either a supercooled liquid, a
structural glass, a plastic crystal, a glassy crystal and an
orientationally ordered crystal within a small temperature
interval around 100 K. Therefore, it has been investigated
experimentally to explore the role of translational degrees of
freedom (TDOF) and ODOF for glassy behavior \cite{RamVie97}. These
experiments have shown that the ODOF of molecular crystals exhibit
quite similar glassy behavior than conventional supercooled
liquids. Additionally, comparing different molecular crystals with each
other, similar glassy behavior was found \cite{RBranLu}.
These similarities also include dynamical heterogeneities
\cite{WinZi03}. The largest deviations of molecular crystals from
supercooled liquids were observed in dielectric spectroscopy. The
former exhibit a rather weak excess wing, or even no such wing, in
contrast to supercooled liquids \cite{RBranLu}.

An interesting model for molecular crystals has been studied some
years ago. The molecules were approximated by infinitely thin hard
rods with length $L$ which were either fixed with their centers on
a fcc-lattice \cite{CRenLoe97} or with their endpoints on a
sc-lattice \cite{SPOb97}. MD- \cite{CRenLoe97} and MC-simulations 
\cite{SPOb97}, respectively, have shown the existence of glasslike 
dynamics. Particularly a critical length $l_c=L/a$ ($a$ is the 
lattice constant) has been determined at which an orientational 
glass transition occurs \cite{CRenLoe97,SPOb97}. However, this transition 
is not sharp, in close analogy to supercooled liquids. The system
of infinitely thin hard rods is particularly interesting since
there are \textit{no} static orientational correlations.
Consequently, glassy behavior does not originate from growing
static correlations, but results from \textit{entanglement} which
leads to a ``dynamical cage''.

As far as we know there is no \textit{microscopic} theory which
describes glassy behavior of molecular crystals with
self-generated disorder. For mixed crystals \cite{UTHoe90},
i.e.~crystals with \textit{quenched disorder}, a microscopic theory
for the glass transition has been worked out \cite{KHMi87}. This
theory is based on MCT and takes into account ODOF and TDOF,
i.e.~lattice displacements, as well as translation-rotation
\cite{RML94} coupling. The displacements are crucial, since the
statistical substitution, of e.g.~CN molecules in KCN by Br atoms,
leading to the well-known mixed crystal compound (KBr)$_{1-x}$
(KCN)$_x$ \cite{UTHoe90}, generates random displacements. Due to
the translation-rotation coupling, these random displacements
induce random fields acting on the ODOF. MCT was also applied to
spin glass models, where the coupling constants between spins are
at random \cite{WGoeSj84}. Unfortunately, the quality of MCT
predictions for mixed crystals and spin glasses has not really
been tested, in contrast to supercooled liquids
\cite{WGoe91,RSRi54,WGoe98,WGoe99,KBin03}.

Since MCT has been very successful
\cite{WGoe91,RSRi54,WGoe98,WGoe99,KBin03} to describe glassy
dynamics of supercooled liquids, and since it has also been
applied to mixed crystals and spin glasses, it is natural to
derive MCT equations for plastic crystals, as well. This will be
done in Sec.~\ref{secII}. The calculation of the glass transition
line and the critical nonergodicity parameters from the MCT equations
will be presented in Sec.~\ref{secIII} for hard ellipsoids on a
sc-lattice. The final section contains a discussion of the
results and some conclusions. More technical details are
put into four appendices.
\section{Theoretical framework}
\label{secII}
In this section we will describe how MCT equations can be derived
for molecular crystals. The strategy is quite similar to that for
simple liquids \cite{WGoe91,RSRi54}, molecular liquids of linear
molecules \cite{RSTS97,RSP02} and arbitrary molecules
\cite{LFa99}. The introduction of the microscopic orientational
density, the corresponding current density and their
time-dependent correlators will be described in subsection~\ref{secIIA}.
Then, in subsection~\ref{secIIB}, we apply the Mori-Zwanzig projection
formalism \cite{JPHa86,DFor75} to derive an equation of motion for
the time-dependent orientational correlators. Following MCT for
liquids, the memory kernel is then approximated by a bilinear
superposition of the time-dependent orientational correlators.
\subsection{Microscopic orientational densities and their correlators}
\label{secIIA}
We consider a Bravais lattice with $N$ lattice sites. Since the
experimental and simulational results for supercooled molecular
crystals \cite{JLSan82,RBranLu,AffDes99,JKoga00,AffCoDe01,ASriBe96,
RamVie97,WinZi03} have demonstrated that glassy behavior can occur 
due to steric hindrance of the ODOF, we restrict ourselves to a 
\textit{rigid} lattice, i.e. we neglect the translation-rotation 
coupling. Then, the increase of steric hindrance by decreasing temperature
can be accounted for by either a variation of the lattice constant
or equivalently by an increase of the size of the molecules. At each
lattice site we fix a molecule. The natural way is to fix its center 
of mass. All molecules are assumed to be identical and rigid, as well. 
We will consider \textit{linear} molecules only. Generalization to 
\textit{arbitrary} molecules can be performed like for molecular 
liquids \cite{LFa99}.

It is also obvious that the ODOF are best described in a
molecule-fixed frame with its origin coinciding with the lattice
site. In principle one could choose any other reference point
\cite{RSP02}. But this would \textit{artificially} introduce TDOF,
besides the ODOF. Using the former choice, 
the ODOF of the $n$-th molecule at the site 
with lattice vector ${\bf x}_n$ is given by the angles $\Omega_n=
(\theta_n, \phi_n)$. The third angle $\chi_n$ with respect to the 
symmetry axis is irrelevant for the glassy dynamics. The moment of 
inertia for the axes perpendicular to the symmetry axis is denoted 
by $I$. The interaction between the molecules is given by 
$V( \Omega_1, \ldots, \Omega_N)$ and the classical dynamics 
follows from the classical Hamiltonian
\begin{align} 
\label{eq1}
H(\{\theta_{n},&\phi_{n}\},\{p_{\theta_{n}},p_{\phi_{n}}\}) =
\nonumber\\
&=\frac{1}{2I}\,\sum\limits_{n =1}^{N}\,\left[ \,p_{\theta_{n}}^{2} +
\frac{p_{\phi_{n}}^{2}}{\sin^{2}\theta_{n}}\,\right] +
V(\{\theta_{n},\phi_{n} \})\,,
\end{align}
where $p_{\theta_{n}}$ and $p_{\phi_{n}}$ are the momenta conjugate to
$\theta_{n}$ and $\phi_{n}$, respectively.

Next we introduce the microscopic, local orientational density
\begin{equation}
\label{eq2}
\rho_{n}(\Omega,t) = \delta(\Omega|\Omega_{n}(t))\,,
\end{equation}
with $\delta(\Omega|\Omega') =(\sin\theta)^{-1} \delta (\theta
-\theta')\delta (\phi -\phi')$ and $\Omega_n(t)$ the classical
trajectory of the $n$-th molecule. The one-molecule orientational
density $\rho^{(1)}(\Omega)$ is given by
\begin{equation}
\label{eq3}
\rho^{(1)}(\Omega) = \big\langle\,\delta(\Omega|\Omega_{n}(t))\,
\big\rangle\,,
\end{equation}
which is independent on $t$ and, for identical molecules, also on 
$n$. $\langle(\cdot)\rangle$ denotes canonical averaging with 
respect to initial conditions in the $4 N$-dimensional phase space.

Taking the time derivative of $\rho_n(\Omega, t)$ leads to 
the continuity equation.
\begin{equation}
\label{eq4} \dot{\rho}_{n}(\Omega,t) = i \,\hat{{\bf L}}_{\Omega_{n}}
\cdot {\bf j}_{n}(\Omega,t)\,.
\end{equation}
Here, $\hat{{\bf L}}_{\Omega_{n}}$ is the angular momentum operator
acting on $\Omega_{n}$, and
\begin{equation}
\label{eq5} {\bf j}_{n}(\Omega,t) = {\boldsymbol \omega}_{n}(t) \,
\delta(\Omega|\Omega_{n}(t)) \equiv {\boldsymbol \omega}_{n}(t)
\,\rho_{n}(\Omega,t)
\end{equation}
is the corresponding orientational current density, which involves
the angular velocity ${\boldsymbol \omega}_n (t)$. We also
introduce the ``longitudinal'' orientational current density
\begin{equation}
\label{eq6} j_{n}(\Omega,t) = \hat{{\bf L}}_{\Omega_{n}} \cdot {\bf
j}_{n}(\Omega,t)\, .
\end{equation}
With these quantities we can define the time-dependent
orientational correlators:
\begin{equation}
\label{eq7} G_{nn'}(\Omega,\Omega',t) = \big\langle
\,\delta\rho_{n}(\Omega,t)\, \delta\rho_{n'}(\Omega') \,\big\rangle
\end{equation}
of the local orientational density fluctuations
\begin{equation}
\label{eq8}
\delta \rho_{n}(\Omega,t) = \rho_{n}(\Omega,t) -
\big\langle \,\rho_{n}(\Omega,t) \,\big\rangle = 
 \rho_{n}(\Omega,t) - \rho^{(1)}(\Omega)\,,
\end{equation}
as well as
\begin{equation}
  \label{eq9}
  J_{nn'}(\Omega,\Omega',t) =
  4\pi \,\big\langle \,j_{n}(\Omega,t) \,
    j_{n'}(\Omega') \,\big\rangle \,.
\end{equation}
Here, $\delta\rho_{n}(\Omega) \equiv \delta\rho_{n}
(\Omega,0)$ and $j_{n}(\Omega)  \equiv j_{n}(\Omega,0)$.

Similar to molecular liquids \cite{RSTS97,RSP02}, we expand
the orientation-dependent functions with respect to spherical
harmonics $Y_\lambda(\Omega)$, $\lambda=(lm)$, as already done for
the static correlators \cite{MRic04}\footnote{Instead, one could 
also use a complete set of functions, determined by the rotational
symmetry. If $\mathcal{P}$ is the point symmetry group of the
lattice and $\mathcal{P}_{M}$ the symmetry group of the molecules,
one can use basis functions for irreducible representations of the
symmetry group of $\rho^{(1)}(\Omega)$, which is a subgroup of
$\mathcal{P} \otimes \mathcal{P}_{M}$~\cite{RML94,YP1980,BP1988}. 
For axially symmetric particles, these are linear combinations of
the spherical harmonics $Y_{\lambda}(\Omega)$, $\lambda = (lm)$
\cite{RML94,YP1980}}. This allows to represent any functions
$f_n(\Omega, t)$ and $F_{nn'}(\Omega,\Omega',t)$ by their $\lambda$-
and Fourier-transforms. The corresponding transform of
$G_{nn'}(\Omega,\Omega',t)$ leads to the intermediate
scattering functions
\begin{equation}
\label{eq10}
S_{\lambda\lambda'}({\bf q},t) = \frac{4\pi}{N}\;
\big\langle\, \delta\rho^{*}_{\lambda}({\bf q},t) \,
\delta\rho_{\lambda'}({\bf q}) \, \big\rangle
\end{equation}
and the corresponding current density correlators
\begin{align}
\label{eq11} J_{\lambda\lambda'}({\bf q},t) &= \frac{4\pi}{N}\;
\big\langle\, j^{*}_{\lambda}({\bf q},t) \, j_{\lambda'}({\bf q}) \,
\big\rangle \,.
\end{align}
These correlators form matrices ${\bf S} ({\bf q}, t)=(S_{\lambda
\lambda'}({\bf q}, t))$, etc. The wave vectors ${\bf q}$ are
restricted to the 1. Brillouin zone (1.BZ), due to the lattice 
translational invariance. The correlators $S_{\lambda \lambda'}
({\bf q},t)$ form a complete set. For example the neutron scattering 
function $S_{\rm neutron} ({\bf q}, t)$ can be expressed by 
$\{S_{\lambda \lambda'}({\bf q},t)\}$ using the scattering 
lengths of the molecular sites \cite{CThRS99}.

Note that the correlators (\ref{eq10}) and (\ref{eq11})
vanish for $l=0$ and/or $l'=0$, due to the absence of TDOF. 
The symmetries of the orientational correlators discussed 
in Ref. \cite{MRic04} also hold for the time dependent quantities.
They will be applied to reduce the number of independent correlators.
\subsection{Mode coupling theory}
\label{secIIB}
The goal of this subsection is to derive an approximate equation
of motion for the intermediate scattering functions $S_{\lambda
\lambda'} ({\bf q}, t)$ of molecular crystals. Due to the rigid
lattice, only ODOF are involved. In case that the steric hindrance is
large enough the orientational density fluctuations $\delta
\rho_\lambda ({\bf q}, t)$ contain slow parts. Choosing
$\delta \rho_\lambda ({\bf q}, t)$ and the ``longitudinal''
current density $j_\lambda ({\bf q}, t)$ as slow variables, we can
apply the Mori-Zwanzig formalism \cite{JPHa86,DFor75} to derive an
equation of motion for ${\bf S}({\bf q}, t)$:
\begin{align}
\label{eq12}
\ddot{{\bf S}}({\bf q},t)&+{\bf J} \, 
{\bf S}^{-1}({\bf q}) \,{\bf S}({\bf q},t) +
\nonumber\\
&+ \int\limits_{0}^{t} \text{d}t' \,{\bf M}({\bf q},t - t') 
\,{\bf J}^{-1} \, \dot{{\bf S}}({\bf q},t') = {\bf 0}\,.
\end{align}
The notation $^{-1}$ means the inverse of the $l,l'>0$ block of the 
respective matrix, i.e. the inverse with respect to the subspace of 
non-constant functions in angular space. This is because the first 
rows and columns of these matrices vanish. The only exception from 
this rule is ${\bf d}^{-1}$ in App.~\ref{appD}. The prefactor ${\bf J} 
\, {\bf S}^{-1} ({\bf q})$ in Eq.~(\ref{eq12}) is related to 
\begin{align}
\label{eq13}
{\bf \Omega}^{2}({\bf q}) &= {\bf S}^{-1/2}({\bf q})\,{\bf J} \, 
{\bf S}^{-1/2}({\bf q})\,,
\end{align}
which is the square of the symmetric microscopic frequency matrix 
$(\Omega_{\lambda\lambda'} ({\bf q}))$. It depends on the static 
orientational correlators ${\bf S} ({\bf q})$ and on ${\bf J} \equiv 
(J_{\lambda \lambda'} ({\bf q}))$, which is independent of ${\bf q}$ 
(see App. \ref{appA}). The matrix elements of the memory kernel ${\bf M}
({\bf q},t)$ are given by
\begin{equation}
\label{eq14}
M_{\lambda\lambda'}({\bf q},t) =
\frac{4\pi}{N} \, \big\langle \left(\mathcal{L}
\, j_{\lambda} ({\bf q}) \right)^{*}  \big| \, Q\,\text{e}^{-i \,
  Q\,\mathcal{L} \, Q \, t} \, Q\,\big| \,\mathcal{L}
\,j_{\lambda'}({\bf q}) \,\big\rangle \,,
\end{equation}
the correlations of the fluctuating forces $Q\,| \,{\cal L}
\,j_\lambda ({\bf q})\,\rangle$. ${\cal L}$ is the Liouville 
operator and $Q=1-P_{\rho}-P_{j}$ (see Eq. (\ref{eqB3})) 
projects perpendicular to the slow variables $\delta
\rho_\lambda ({\bf q})$ and $j_\lambda ({\bf q})$. 

In a final step we perform the mode coupling approximation for the
slow part ${\bf m}({\bf q},t)$ of ${\bf J}^{-1}\,{\bf M}({\bf q},t)\,
{\bf J}^{-1}$, which enters Eq.~(\ref{eq23}), yielding (see
Appendices \ref{appB}-\ref{appD})
\begin{align}
\label{eq15} 
m_{\lambda\lambda'}&({\bf q},t)  \approx \frac{1}{2N}
\, \sum\limits_{{\bf Q}} \,\sum\limits_{{\bf q}_{1},{\bf q}_{2}
\atop \in \text{1.BZ}}\!'\sum\limits_{\lambda_{1}\lambda_{1}'
\lambda_{2}\lambda_{2}'}\hspace{-2ex}'\\[1ex]
& V({\bf q}\lambda\lambda'|{\bf q}_{1}\lambda_{1}\lambda_{1}'; {\bf
q}_{2}\lambda_{2}\lambda_{2}')\, S_{\lambda_{1}\lambda_{1}'}({\bf
q}_{1},t) \, S_{\lambda_{2}\lambda_{2}'}({\bf q}_{2},t) \,. \nonumber
\end{align}
The vertices are
\begin{align}
\label{eq16}
&V({\bf q}\lambda\lambda'|{\bf
q}_{1}\lambda_{1}\lambda_{1}'; {\bf q}_{2}\lambda_{2}\lambda_{2}')=
\nonumber\\[2ex]
&= \frac{1}{(4\pi)^{2}} \,\sum\limits_{\lambda_{3}\lambda_{3}'}\!'\; 
\big(\widetilde{\bf J}^{-1}\big)_{\lambda\lambda_{3}} 
\left[\,\sum\limits_{\lambda''}\!'\, \, v({\bf q}\lambda_{3}|{\bf
q}_{1}\lambda_{1}; {\bf q}_{2}\lambda_{2};\lambda'') \,\right]
\times
\nonumber\\[2ex]
&\times\left[ \,\sum\limits_{\lambda'''}\!'\, \, v({\bf
q}\lambda_{3}'|{\bf q}_{1}\lambda_{1}'; {\bf
q}_{2}\lambda_{2}';\lambda''') \,\right]^{*} \big(\widetilde{\bf
J}^{-1}\big)_{\lambda_{3}'\lambda'}\;,
\end{align}
where
\begin{align}
\label{eq17}
&v({\bf q}\lambda|{\bf q}_{1}\lambda_{1};
{\bf q}_{2}\lambda_{2};\lambda'') =
\nonumber \\[3ex]
&= b(l''l_{2}l)\,C(l''l_{2}l,m''m_{2}m)\,
c_{\lambda''\lambda_{1}}({\bf q}_{1}) +
(1 \leftrightarrow 2)\,,
\end{align}
\begin{align}
\label{eq18}
&\,b(ll'l'') = \frac{1}{2}\,i^{l+l'-l''} \left[ \frac{(2l+1)(2l'+1)}
{2l''+1} \right]^{\frac{1}{2}} \times\\[3ex]
&\times\left[ \, 1 + (-1)^{l+l'+l''}\, \right] 
\sqrt{l(l+1)} \, \sqrt{l''(l''+1)}\; C(ll'l'',101)\,,\nonumber
\end{align}
and
\begin{equation}
\label{eq19}
\widetilde{J}_{\lambda\lambda'} = \frac{I}{kT}\,J_{\lambda\lambda'} \,,
\end{equation}
is the inertia and temperature-independent part of $J_{\lambda 
\lambda'}$. $\sum_{{\bf q}_1,{\bf q}_2}\!'\,$ denotes summation 
such that ${\bf q}_1 + {\bf q}_2= {\bf q} + {\bf Q}$, with 
${\bf Q}$ a reciprocal lattice vector, and $\sum_{\lambda}\!'\,$ 
indicates summation over all $\lambda \neq (00)$. $C(l l' l'', 
m m'm'')$ are the Clebsch-Gordon coefficients and $c_{\lambda
\lambda'} ({\bf q})$ the direct correlation function matrix 
elements. 

The result, Eqs.~(\ref{eq15})-(\ref{eq19}), has a striking
similarity to the slow rotational part $m^{\rm RR}_{\lambda 
\lambda'}({\bf q}, t)$ of the memory kernel for molecular liquids
\cite{RSTS97,RSP02}. This is not surprising. Particularly
$\upsilon({\bf q} \lambda | {\bf q}_1 \lambda_1; {\bf q}_2
{\lambda_2}; \lambda'')$ are identical, up to a factor $\sqrt{l (l
+1)}$. This similarity originates from the factorization of a 
static three-point correlator described in Appendix \ref{appD}.
It is this approximation which leads to the rather simple result,
Eqs. (\ref{eq16}) and (\ref{eq17}), for the vertices. Of course,
taking the static three-point correlator from a simulation would
make this factorization approximation unnecessary. However, we do
not expect any qualitative influence using our approximation instead
of the correct simulational result. Such an influence is only to be
expected for systems which interact through three-, four- $\ldots$
body potentials. Since the molecules are fixed on lattice sites,
such covalent bonds may be less important for molecular crystals.

There are four main differences with
respect to MCT for molecular liquids. First,
the tensorial MCT equations for molecular liquids are first order
integro-differential equations which can not be transformed to a
second order integro-differential equation, like Eq.~(\ref{eq12}). 
Second, for molecular crystals  $m_{\lambda \lambda'} ({\bf q}, t)$
contains a sum over reciprocal lattice vectors ${\bf Q}$. Therefore,
the sum over ${\bf q}_1, {\bf q}_2$ involves \textit{Umklapp} processes.
Third, due to the rigid lattice, only the $l,l'>0$ matrix elements are 
nonzero. Fourth, the static current density correlator $J_{\lambda
\lambda'}$ in Eq.~(\ref{eq15}) does not cancel completely, as it does 
for a molecular liquid. There remains an anisotropic part $\tilde{J}_
{\lambda \lambda'}$, which equals $l(l + 1)\delta_{\lambda\lambda'}$ 
for a liquid and is defined in Eq.~(\ref{eq19}). $\tilde{J}_{\lambda
\lambda'}$, can be related to the $\lambda$-transform $\rho^{_{(1)}}_
{_\lambda}$ of the one-molecule orientational density
$\rho^{(1)}(\Omega)$, which is needed as an input 
(for details see App.~\ref{appA}).

There is no explicit dependence 
of the kernel ${\bf m}({\bf q},t)$ on $T$ and $I$. On a
large time scale we can neglect $\ddot{{\bf S}}({\bf q}, t)$ in
Eq.~(\ref{eq12}). The remaining equation does not involve any 
inertia effect, i.e. the glassy dynamics depends not on $I$, 
except for fixing the time scale.

The vertices $V({\bf q} \lambda \lambda'| {\bf q}_1 \lambda_1
\lambda'_1; {\bf q}_2 \lambda_2 \lambda'_2)$ depend on
$\tilde{{\bf J}}$ and the direct correlation function 
${\bf c}({\bf q})$ only, where ${\bf c}({\bf q})$ is related to the 
\textit{static} orientational correlators ${\bf S}({\bf q})$ 
by the Ornstein-Zernike (OZ) equation for molecular crystals 
\cite{MRic04}:
\begin{equation}
\label{eq20} {\bf S}({\bf q}) = \left[{\bf D}^{-1} -
  \frac{1}{4\pi} \, {\bf c}({\bf q}) \right]^{-1} .
\end{equation}
This equation is similar to that for molecular liquids
\cite{RSTS97,CGG84}\footnote{The static orientational correlators 
written in complete analogy to simple and molecular liquids are 
${\bf S}({\bf q}) = \big[\,{\bf 1} - \widetilde{{\bf D}}^{\frac{1}{2}} 
\, {\bf c}({\bf q})\,\widetilde{{\bf D}}^{\frac{1}{2}} \big]^{-1}$, 
$\widetilde{{\bf D}} = \frac{1}{4\pi} {\bf D}$. This form is 
equivalent to Eq.~(\ref{eq20}), but not so handy.}, with 
exception of the appearance of ${\bf D}$. $D_{\lambda\lambda'}$ 
is the $\lambda$-transform of $D(\Omega, \Omega')=4\pi\,
[\,\rho^{(1)}(\Omega) \,\delta(\Omega|\Omega') - \rho^{(1)}(\Omega) 
\,\rho^{(1)}(\Omega')\,]$ and can be expressed by the 
$\lambda$-transform of $\rho^{(1)} (\Omega)$, too.

This discussion makes clear that the closed set of MCT equations
(\ref{eq12})-(\ref{eq19}) requires two different \textit{static}
input quantities, the one-molecule quantities $D_{\lambda
\lambda'}, \tilde{J}_{\lambda \lambda'}$ which can be expressed by
$\{ \rho^{(1)}_\lambda \}$ and the two-molecule correlators
$S_{\lambda \lambda '} ({\bf q})$, or $c_{\lambda \lambda'} ({\bf
q})$. Note also that we have neglected contributions to
$m_{\lambda \lambda'} ({\bf q}, t)$ coming from the fast part of
$\delta \rho ({\bf q}, t)$ which leads to a damping term in
Eq.~(\ref{eq12}). On a long time scale this has no influence.

Here we restrict ourselves to the investigation of 
the orientational glass transition itself. For this we 
introduce the nonergodicity parameters (NEP, not normalized!)
\begin{equation}
\label{eq21} F_{\lambda\lambda'}({\bf q}) = \lim_{t \to \infty}\,
S_{\lambda\lambda'}({\bf q},t) \, .
\end{equation}
In the limit $t \to \infty$, Eq.~(\ref{eq12}) leads to
\begin{equation}
\label{eq22}
{\bf S}^{-1}({\bf q})\,{\bf F}({\bf q}) \, 
\left[\,{\bf S}({\bf q}) - {\bf F}({\bf q})\,
\right]^{-1} = \mathcal{F}[{\bf F}({\bf q})]
\end{equation}
where
\begin{equation}
\label{eq23}
 \mathcal{F}[{\bf F}({\bf q})] = \lim_{t \to \infty}\,{\bf m}({\bf q},t)\,.
\end{equation}
Eqs.~(\ref{eq22}) and (\ref{eq23}) are matrix equations for $l,l' >0$,
since the first columns and rows of the involved matrices vanish,
due to the rigid lattice.
\section{Results for hard ellipsoids}
\label{secIII}
After having derived equations of motion for the orientational
correlators $S_{\lambda\lambda'}({\bf q},t)$, their time dependence
could be calculated numerically. Although the mathematical structure
of the MCT-equations (\ref{eq12}), (\ref{eq15})-(\ref{eq19}) is 
identical to that of more-component liquids, the numerical solution
is hampered due to the \textit{anisotropy} of the lattice, in
contrast to liquids. Because of this anisotropy the correlators 
also depend on the direction of ${\bf q}$. For liquids it has 
turned out that the restriction to several hundreds of values
for $q = |{\bf q}|$ leads to rather precise solutions of the
MCT-equations. We will see below that we have to choose several 
thousands of ${\bf q}$ vectors within the first Brillouin zone.
In addition, in comparison to molecular liquids, there are 
more independent correlators for each pair $(l,l')$, increasing 
the number of equations even more. Consequently, a numerical 
solution will require either an improvement of the numerical code
\cite{Hofack91} usually used for the numerical solution of the
MCT-equations and/or further simplification of $S_{\lambda\lambda'}
({\bf q})$, e.g. neglecting the dependence on ${\bf q}/|{\bf q}|$.
Since we want to avoid such type of approximations we will restrict
ourselves to the calculation of the glass transition point and
the corresponding critical nonergodicity parameters and leave
the solution of the time-dependent equations for future.
Nevertheless, their identical structure to that of liquids
already ensures the validity of, e.g., the two time scaling
laws \cite{WGoe91} for molecular crystals as well.

In order to solve Eq.~(\ref{eq22}) we have chosen hard ellipsoids
of revolution, fixed with their centers on a sc-lattice with lattice 
constant equal to one. The symmetry axis of the ellipsoids has length $a$ 
and the length of the perpendicular axes is $b$. Replacing linear, 
rigid molecules by hard ellipsoids is probably not a bad approximation 
since the steric hindrance is qualitatively the same. In addition, 
the choice of hard ellipsoids has two advantages: First, we have 
already calculated the static orientational correlators $S_{\lambda 
\lambda'} ({\bf q})$ for $l,l'\leq 4$ within the Percus-Yevick (PY) 
approximation, and $\rho^{(1)}_\lambda$ (which yields $\tilde{J}_
{\lambda \lambda'}$ and $D_{\lambda \lambda'}$) for $l \leq 8$ by 
MC-simulations. Second, we have recently solved the MCT-equations
for a molecular \textit{liquid} of hard ellipsoids \cite{MLeRS00}
which allows to compare the conditions for the appearance of the
ideal glass transition for the ellipsoids on a lattice and in their
liquid phase. This comparison will allow to estimate the qualitative
or quantitative role of TDOF for the freezing of the ODOF. Furthermore,
the ellipsoids' head-tail symmetry leads to a decomposition of
Eq.~({\ref{eq12}) and therefore of Eq.~(\ref{eq22}) into a closed
set of equations for $S_{\lambda \lambda'} ({\bf q}, t)$ and
$F_{\lambda \lambda'} ({\bf q})$, respectively, for $l,l'$
\textit{both even} and a set of these quantities for $l,l'$
\textit{both odd}. All correlators with $l$ even and $l'$ odd or 
vice versa are zero. The set of equations for $l,l'$ both even is 
closed because the memory kernel only contains correlators with $l$ 
and $l'$ even. This is in contrast to the equations for $l,l'$ odd.
The corresponding memory kernel contains a bilinear coupling of
correlators with $l,l'$ even with those where $l, l'$ are odd. 
It is easy to prove that
\begin{equation}
\label{eq24}
S^{(s)}_{\lambda\lambda'}(t) \equiv S_{\lambda\lambda'}({\bf q},t)\,,
\end{equation}
for $l,l'$ both odd. The ``self'' correlator $S^{(s)}_{\lambda
\lambda'} (t)$ is the $\lambda$-transform of $G_{nn}(\Omega,
\Omega', t)$, up to a prefactor. In contrast to this the ``self''
correlator with $l,l'$ even is given by
\begin{equation}
\label{eq25} S^{(s)}_{\lambda\lambda'}(t) = \frac{1}{N}\,
\sum_{{\bf q} \in \text{1.BZ}}\, S_{\lambda\lambda'}({\bf q},t)\,.
\end{equation}
Similar relations hold for the NEP.

There is a technical disadvantage connected with the hard
body potential between the ellipsoids. The inversion of the static
correlators, occurring e.g. in the projectors (cf.~Eq.(\ref{eq14}))
and vertices (cf.~Eq.~(\ref{eq16})), needs some caution. Readers
interested in this point are referred to Refs. \cite{MRic04,MRicker04}.
We stress that the second paper cited in Ref. \cite{MRic04} and 
\cite{MRicker04} contain additional technical information,
particularly the discussion of those mathematical problems related
to hard core interactions. However, these details will not be
needed in the following.

The numerical solution of Eqs.~(\ref{eq22}),~(\ref{eq23}) requires 
a restriction of the matrices to $l,l' \le l_{\rm  max}$. We
have chosen $l_{\rm max}=4$. The ${\bf q}$-vectors are discretized,
i.e. for the $\alpha$-component of ${\bf q}$ we have chosen $q_\alpha=
\nu_{\alpha}\frac{2 \pi}{M}$, $\nu_\alpha =-\frac{M}{2},-\frac{M}{2}+1,
\cdots,0,\cdots, \frac{M}{2}-1$ with $M=32$, which makes a total of $32768$
${\bf q}$-vectors. Due to the point symmetry of the lattice, the number of
independent $F_{\lambda\lambda'}({\bf q})$ can be reduced. For more
details the reader is referred to Refs. \cite{MRic04,MRicker04}. The 
solution of Eqs.~(\ref{eq22}) yields the NEP 
$F_{\lambda \lambda'}({\bf q})$ and the corresponding normalized
quantities $f_{\lambda \lambda'}({\bf q})$, where the normalization 
$f_{\lambda \lambda'}({\bf q})= F_{\lambda \lambda'} ({\bf q})\,[\,
S_{\lambda\lambda}({\bf q})\,S_{\lambda'\lambda'}({\bf q})\,]^{-1/2}$ 
has been chosen. Varying the aspect ratio $X_0=\frac{a}{b}$ and the 
volume fraction $\varphi =\frac{\pi}{6} ab^2$, we have located the 
glass transition line, at which for $l,l'$ both even a nontrivial 
solution for $F_{\lambda \lambda'} ({\bf q})$ bifurcates. 

\begin{figure}[h]
\vspace{-0.2cm}
\includegraphics[width=5.5cm,angle =270]{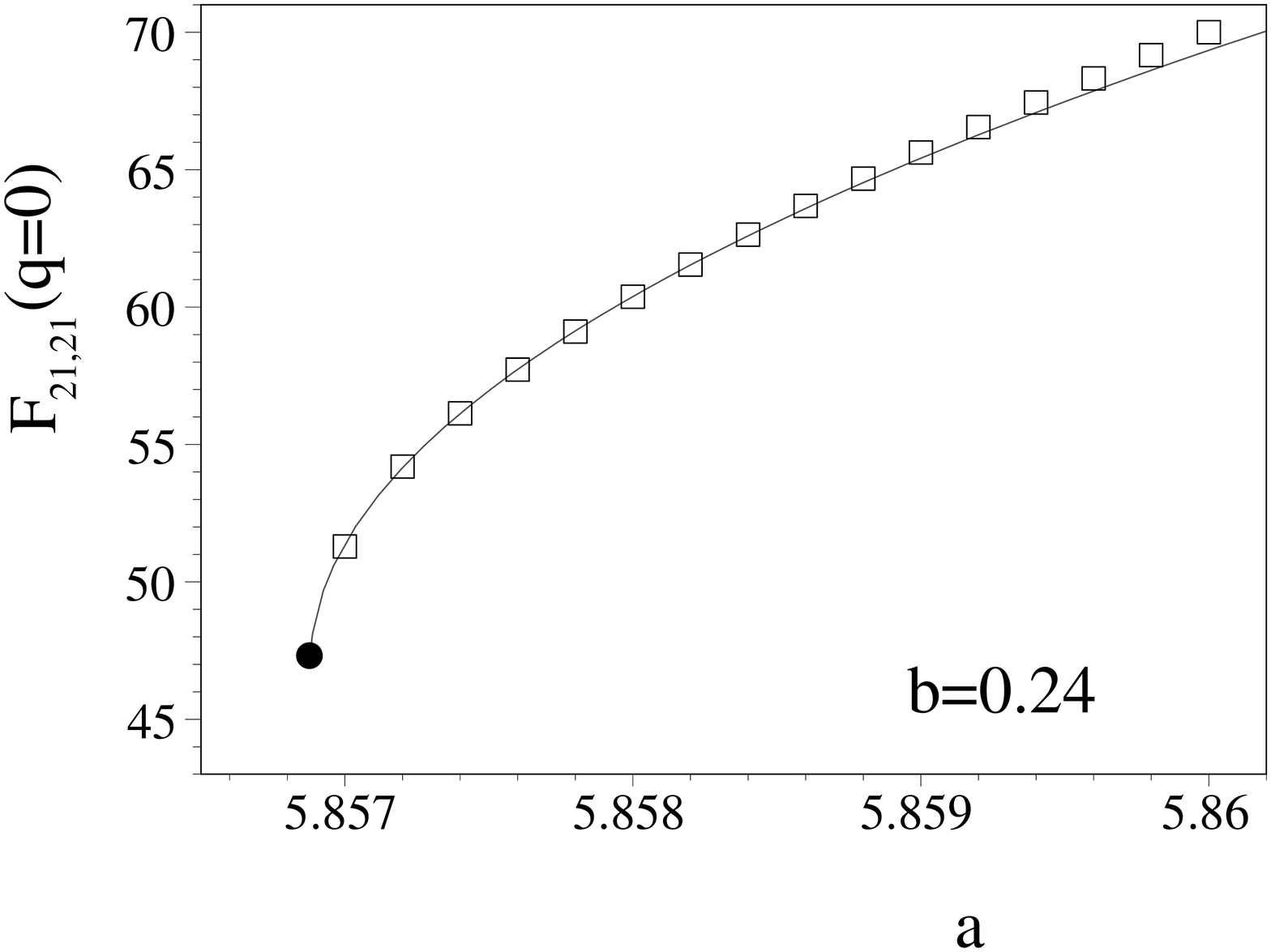}
\vspace{-0.2cm}
\caption{\label{fig1}$a$-dependence of the NEP $F_{21,21}({\bf q} =
  {\bf 0})$ in the vicinity of the glass transition point for fixed 
  $b = 0.24$ and different values of $a$ (squares). Also shown is the 
  square root fit $47 + 400 \sqrt{a-5.85688}$, leading to the
  critical values $(a_{c},F_{21,21}^{c}({\bf 0}))$ (black dot). Note
  the rather large prefactor of the square root.}
\end{figure}
This has been done by approaching the glass transition point from the 
glass side. Fig.~\ref{fig1} shows an example where $F_{21,21}({\bf q}
= {\bf 0})$ is represented as function of $a$ for fixed $b$. The fit
with a square root, predicted by MCT for a type B-transition 
\cite{WGoe91} allows to locate the glass transition point $a_{c}(b)$
for fixed $b=0.24$ up to a relative deviation better than
$10^{-4}$! The NEP $F_{21,21}({\bf 0})$ for $a=5.857$ (see
Fig.~\ref{fig10}) deviates less than ten percent
from the critical NEP $F^{c}_{21,21}({\bf 0})= 47$ at
$a_{c}=5.85688$, and no qualitative change is to be expected on
further approach towards $a_{c}$. 
\subsection{Phase diagram }
\begin{figure}[ht]
\includegraphics[width=8.1cm,angle =0]{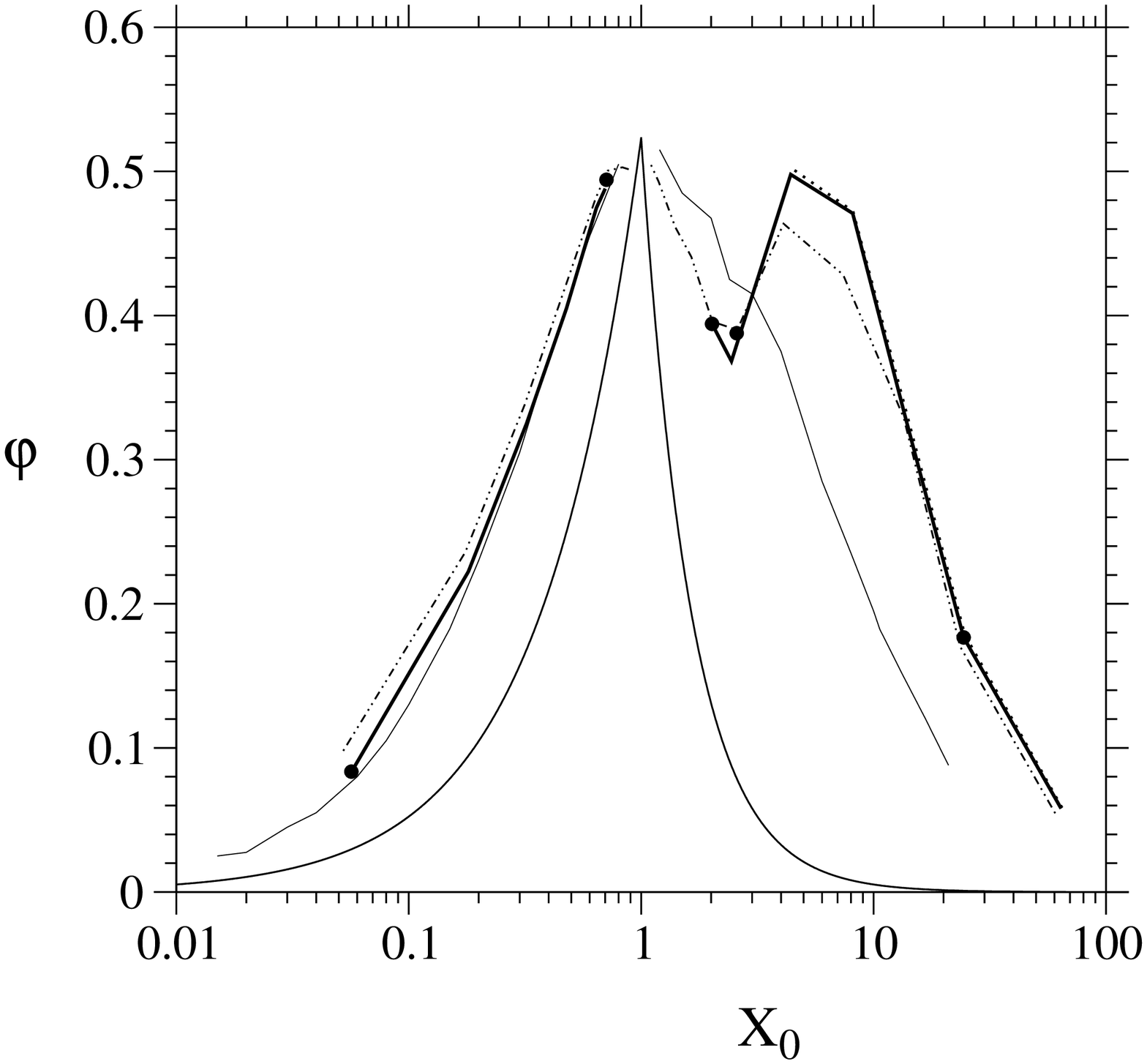}
\vspace{-4.1cm}
\caption{\label{fig2}Phase diagram of hard ellipsoids on a sc 
  lattice. Shown is the curve below which the ellipsoids are free 
  rotators (full line with cusp at $X_{0} =1$), the equilibrium phase
  transition line $\varphi_{\text{eq}}(X_{0})$ from MC simulations 
  (full thin lines), the line $\varphi_{\text{PY}}(X_{0})$ of highest
  densities to be reached by numerical solution of the OZ/PY equations
  (dash-dotted lines), the line $\varphi_{\text{extra}}(X_{0})$ for
  $X_{0}\gtrsim 4$, where the extrapolated OZ/PY results diverge
  (dotted line) and the MCT glass transition line $\varphi_{c}(X_{0})$ 
  (full thick lines). For $X_{0} \gtrsim 8$, $\varphi_{\text{extra}}
  (X_{0})$ and $\varphi_{c}(X_{0})$ are almost identical. $\bullet$
  denote the five state points for which the nonergodicity parameters
  presented in Figs. \ref{fig3}-\ref{fig12} were calculated.}
\end{figure}
The glass transition line $\varphi_{c}(X_{0})$
obtained in this way is shown in Fig.~\ref{fig2}.
Fig.~\ref{fig2} also contains the equilibrium phase transition line
$\varphi_{\text{eq}}(X_{0})$ from MC-simulations, and the line
$\varphi_{\text{PY}}(X_{0})$. Also shown is the curve 
$\varphi_{\text{extra}}(X_{0})$ where the extrapolated OZ/PY static
orientational correlators for $X_{0} \gtrsim 4$ at the zone center
diverge. $\varphi_{\text{eq}}(X_{0})$ and $\varphi_{\text{PY}}(X_{0})$
were obtained from the corresponding lines $a_{\text{eq}}(b)$ and
$a_{\text{PY}}(b)$ of Ref. \cite{MRic04}. Finally, the
solid line with the cusp at $X_{0} = 1$ is the location of all
$\varphi(X_{0})$ at which the rotators start to interact.
 
At $\varphi_{\text{eq}}(X_{0})$, an equilibrium phase transition 
from a (dynamically) disordered to an orientationally ordered phase
occurs. The line $\varphi_{\text{PY}}(X_{0})$ locates the $(X_{0},
\varphi)$-pairs for which the iterative numerical procedure to solve 
the OZ/PY equations becomes unstable. This is associated with some of
the maxima of ${\bf S}({\bf q})$ becoming very large, giving evidence
of a divergency. Since the $X_{0}$-dependence of $\varphi_\text{eq}$
and $\varphi_\text{PY}$ is qualitatively similar, this behavior may
indicate an equilibrium phase transition, as speculated for a liquid 
of hard ellipsoids \cite{MLe99}. However, in contrast to the latter, 
the deviation of $\varphi_{\text{PY}}(X_{0})$ from $\varphi_{\text{eq}}
(X_{0})$ is much larger, especially for prolate ellipsoids. Using the
static correlators from OZ/PY-theory as an input for the calculation
of the NEP from Eqs. (\ref{eq22}) and (\ref{eq23}) we have found
a glass transition for $X_{0} \lesssim 0.7$ and $2 \lesssim X_{0}
\lesssim 2.5$, only. For $0.7 \lesssim X_{0} \lesssim 2$ and $X_{0}
\gtrsim 4$ the system is ergodic for all $\varphi \le 
\varphi_{\text{PY}}(X_{0})$, because the static correlators at the
Brillouin zone center and/or edge are too small. Unfortunately, the
iterative procedure to solve the OZ/PY equations for $0.7 \lesssim
X_{0} \lesssim 2$ and $X_{0} \gtrsim 4$ becomes unstable for $\varphi
\ge \varphi_{\text{PY}}(X_{0})$. Therefore, we have decided to
extrapolate the static correlators to $\varphi \ge \varphi_
{\text{PY}}(X_{0})$. This extrapolation is guided by the physical
assumption that long range orientational order should occur at the
line $\varphi_{\text{extra}}(X_{0})$. It only works
for $X_{0} \gtrsim 4$, but not for the gap in between $X_{0} \approx 0.7$
and $X_{0} \approx 2$. Accordingly, the missing glass transition line
for $0.7 \lesssim X_{0} \lesssim 2$ first of all is based upon the lack
of the static input. Since the ergodic and nonergodic phase are
separated by a critical line of Type-B transitions 
\cite{WGoe91,RSRi54}, $\varphi_{\text{c}}
(X_{0})$ can not terminate at $X_{0} \approx 0.7$ or $X_{0} \approx
2$. There exist two possible scenarios for $\varphi_{\text{c}}(X_{0})$
within this gap. First, $\varphi_{\text{c}}(X_{0})$ converges to 
$\varphi_{\text{c}}(X_{0}=1)$ from above and below $X_{0} = 1$, 
with a possible cusp at $X_{0}=1$. Second, $\varphi_{\text{c}}(X_{0})
\to \varphi_{\text{max}}(X_{0}^{\pm})$ for $X_{0} \to X_{0}^{\pm}$
with $0.7 \lesssim X_{0}^{-} < 1$ and $1 < X_{0}^{+} \lesssim 2$, 
where $\varphi_{\text{max}}(X_{0})$ is the maximum possible volume
fraction of an orientationally disordered configuration
for given $X_{0}$. The second scenario would imply that there
is no glass transition for $X_{0}^{-} \le X_{0} \le X_{0}^{+}$,
i.e. for ellipsoids which are not sufficiently aspherical.

The non-monotonous behavior 
of $\varphi_{\text{PY}}(X_{0})$ for prolate ellipsoids with $1 < X_0 
\lesssim 4 $, which induces a non-monotonicity of $\varphi_c(X_{0})$, 
seems to be an artefact of the PY approximation, as our MC results for
hard prolate ellipsoids suggest, though the static orientational 
correlators from OZ/PY theory are \textit{qualitatively} correct, 
anyway \cite{MRic04}.

If it is true that the divergence of the PY solutions corresponds 
to an equilibrium phase transition, this implies that the ideal 
glass transition is driven by the growth of some $S_{\lambda\lambda'}
({\bf q})$ at the zone center or/and edge due to the growth of the 
orientational order, as will be seen in the following figures. This 
is quite similar to the central peak phenomenon above the equilibrium 
transition temperature at structural phase transitions of first and 
second order \cite{Bru90}. The central peak can be interpreted as a 
quasi-nonergodic behavior and has also been described by MCT \cite{VLA87}. 

The freezing of the $l,l'$ \textit{odd} correlators 
occurs beyond the $l,l'$ even glass transition line and is
treated in subsection~\ref{SecIIID}.
\subsection{Critical nonergodicity parameters}
The critical NEP $F^{c}_{\lambda\lambda'}
({\bf q})$ and the normalized critical NEP 
$f^{c}_{\lambda\lambda'}({\bf q})$ together with the static
orientational correlators are shown in Figs.~\ref{fig3},
\ref{fig6},~\ref{fig7},~\ref{fig9} and~\ref{fig10} for oblate and 
prolate ellipsoids, respectively, along the three highly symmetric 
directions in reciprocal space from the zone center to its 
edge. For each of the three directions and each matrix 
element, a separate subfigure is provided, where the
indices $lml'm'$ are displayed at the top of each figure. 
We have restricted our illustrations to the diagonal
elements $l=l'=2$, $m = m'= 0,1,2$ and 
$l=l'=4$, $m = m'= 0,1,2,3,4$, and the off-diagonal elements 
$l=2$, $l'=4$, $m = m'= 0,1,2$. By the symmetries of the cubic 
lattice, these correlators are all real. 
The scales on the l.h.s. of each tableau belong to 
$S_{\lambda\lambda'} ({\bf q})$ and $F^{c}_{\lambda \lambda'}({\bf q})$, 
those on the r.h.s. to $f^{c}_{\lambda\lambda'}({\bf q})$. Note 
the different scales of the axes for different values of $m=m'$.

For two pairs $(a,b)$ we also present the corresponding
tensorial quantities in real space. Figs.~\ref{fig4} and~\ref{fig8} 
show log-lin representations of the direct space static orientational
correlators $G_{xyz,\lambda\lambda'}$ and the corresponding NEP
$F^{c}_{xyz,\lambda\lambda'}=\lim_{t \to \infty}G_{xyz,\lambda\lambda'}
(t)$ along lattice directions of high symmetry, i.e. $xyz =
00n$, $0nn$ and $nnn$ for $n =0 ,1,\ldots, 8$. Along these directions, all
$G_{xyz,\lambda\lambda'}$ and $F^{c}_{xyz,\lambda\lambda'}$ are real,
too, for $\lambda\lambda'$ as above. Note that a step $\Delta n =1$ 
corresponds to different lengths in direct space, namely $1$, 
$\sqrt{2}$ and $\sqrt{3}$ for the different lattice directions. 
For each $m = m'$ and each lattice direction, a separate figure 
is provided and a logarithmic plotting has been chosen for positive 
and negative values of $G_{xyz,\lambda\lambda'}$ and $F^{c}_{xyz,\lambda
\lambda'}$ separately, i.e. the negative values are presented as 
$-\ln\,|G_{xyz,\lambda\lambda'}|$ and $-\ln\,|F^{c}_{xyz,\lambda\lambda'}|$, 
respectively. The values of $xyz,lml'm'$ are included in each subfigure. 

\begin{figure}[p]
\includegraphics[width=7.5cm,angle =270]{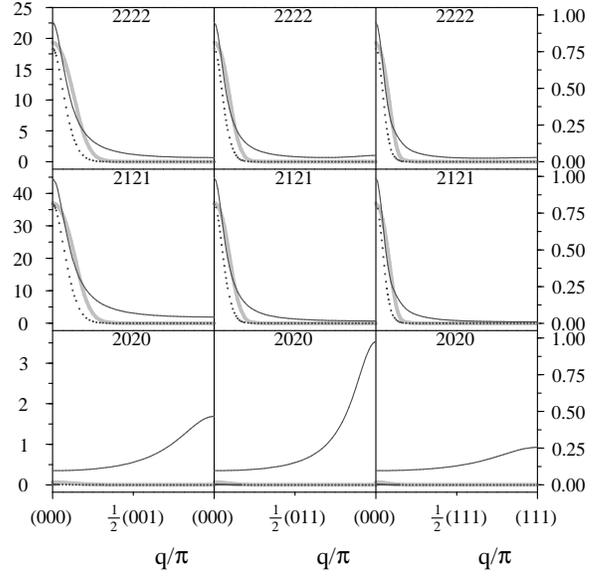}
\caption{\label{fig3}${\bf q}$-dependence of the nonergodicity parameters
  $F^{c}_{\lambda\lambda'}({\bf q})$ (dotted lines), the normalized ones 
  $f^{c}_{\lambda\lambda'}({\bf q})= F^{c}_{\lambda\lambda'}({\bf q})\,
  [\,S_{\lambda\lambda}({\bf q})\,S_{\lambda'\lambda'}({\bf q})\,]^{-1/2}$ 
  (thick grey lines) and of the static structure factors 
  $S_{\lambda\lambda'}({\bf q})$ (solid lines) for $l = l'=2$ and 
  $m =m'= 0,1,2$. Results are shown within the first Brillouin 
  zone along the three highly symmetric reciprocal space directions 
  for \textit{oblate} ellipsoids with $a = 0.08$ and $b = 1.412$, i.e.
  $(X_{0},\varphi) \protect\cong (0.0567,0.0835)$.}
\end{figure}
\begin{figure}[p]
\includegraphics[width=7.5cm,angle =270]{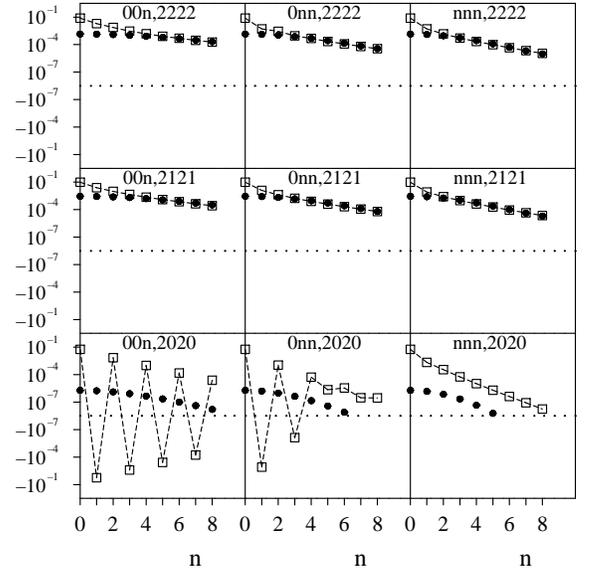}
\caption{\label{fig4}The nonergodicity parameters $F^{c}_{xyz,\lambda\lambda'}$ 
in real space (solid circles) and the static orientational density correlators
$G_{xyz,\lambda\lambda'}$ (squares; dashed lines are a guide to the eye) 
along the three highly symmetric direct lattice directions for \textit{oblate} 
ellipsoids with $a = 0.08$ and $b = 1.412$, i.e. $(X_{0},\varphi)\protect\cong 
(0.0567,0.0835)$. Shown are the diagonal correlators for $l = l'=2$, 
$m =m'= 0,1,2$ and $xyz = 00n$, $0nn$ or $nnn$ for $n =0 ,1,\ldots, 8$.}
\end{figure}
\begin{figure}[h]
\includegraphics[width=5.0cm,angle =270]{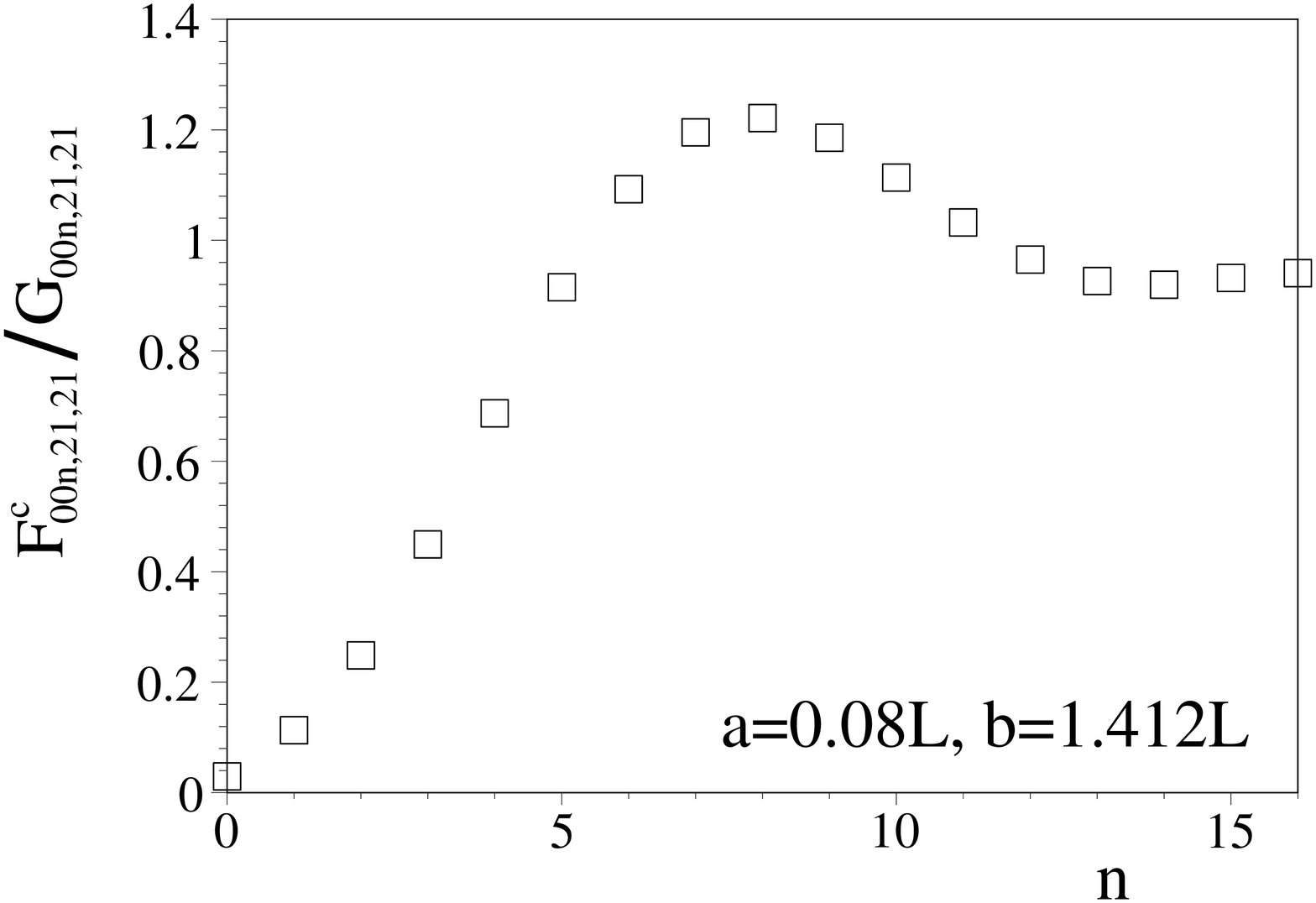}
\caption{\label{fig5}$n$-dependence of the ratio $F^{c}_{00n,21,21}/
  G_{00n,21,21}$ of the NEP $F^{c}_{00n,21,21}$ and his
  static counterpart $G_{00n,21,21}$ for \textit{oblate} ellipsoids 
  with $a = 0.08$ and $b = 1.412$, i.e. $(X_{0},\varphi) \protect\cong
  (0.0567,0.0835)$.}
\end{figure}
\begin{figure}[h]
\vspace{0.5cm}
\includegraphics[width=7.5cm,angle =270]{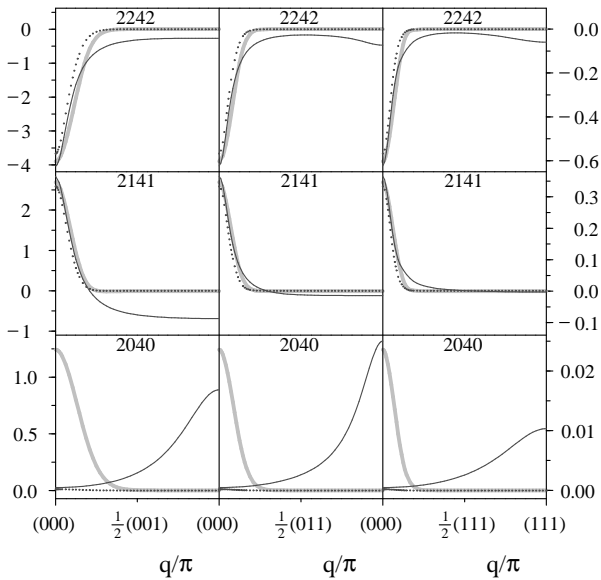}
\caption{\label{fig6}Same as Fig.~\ref{fig3}, but for $l=2$, $l'=4$.}
\end{figure}
Figs.~\ref{fig3}-\ref{fig6} present the NEP for $l=l'=2$ and $l'=2,4$ for 
\textit{oblate} ellipsoids with $a = 0.08$ and $b = 1.412$, which yields 
$(X_{0},\varphi) \protect\cong (0.0567,0.0835)$. In comparison to liquids,
the NEP possess less structure in $q$-space. For $l=l'=2$, they are 
maximum exclusively at the zone \textit{center}. A similar behavior
is found for $l \neq l'$ (cf. Fig.~\ref{fig6}), but here, e.g. for
$m=m'=2$, minima appear instead of maxima. None of the maxima of the 
static structure 
factors $S_{2m,2m}({\bf q})$ and maxima/minima of $S_{2m,4m}({\bf q})$ 
at the zone \textit{boundary} persits in the limit $t \to \infty$. Since 
these maxima belong to alternating orientational density fluctuations,
this proves that such alternating local arrangements 
of the particles do not arrest. This can also be seen in real space.
Fig.~\ref{fig4} exhibits the static orientational density correlators 
$G_{xyz,2m,2m}$ and the $F^{c}_{xyz,2m,2m}$ for $a = 0.08$ 
and $b= 1.412$. Indeed, the oscillations in the correlators
$G_{00n,20,20}$ and $G_{0nn,20,20}$ vanish completely in the long
time limit, while the monotonous decay with $n$ of the $(m=m'>0)$-quantities
is rather stable, even for infinite time. The almost vanishing of
some critical NEP, however, does not require oscillations in the
corresponding $G_{xyz,\lambda\lambda'}$, as can be seen from 
$G_{nnn,20,20}$ and $F^{c}_{nnn,20,20}$.
Another remarkable feature is the behavior at small $n$,
particularly at $n = 0$. Fig.~\ref{fig4} demonstrates that, 
e.g., the magnitude of $F^{c}_{000,2m,2m}$ for $m = 1,2$
is only a few percent or even less of that of $G_{000,2m,2m}$.
Fig.~\ref{fig5} shows that the ratio $F^{c}_{00n,21,21}/G_{00n,21,21}$ 
becomes very small as $n$ is lowered. A similar behavior has been
found for all values $(X_{0},\varphi)$ on the glass transition lines
we have investigated. The dips in $F^{c}_{xyz,\lambda\lambda'}/G_{xyz,
\lambda\lambda'}$ at $n=0$ demonstrate that the relaxation of 
the ``self'' part of the orientational correlators is practically 
not arrested by an orientational cage.

Moving for oblate ellipsoids along the glass transition line 
towards the spherical limit $X_{0} =1$, no qualitatively new 
behavior of the critical NEP is found, but it resembles always 
the characteristics of Figs.~\ref{fig3} and \ref{fig6}. However,
this picture will change as we turn for oblate ellipsoids into 
the glass phase, as will be seen in subsection~\ref{SecIIIC}.

\begin{figure}[ht]
\includegraphics[width=7.5cm,angle =270]{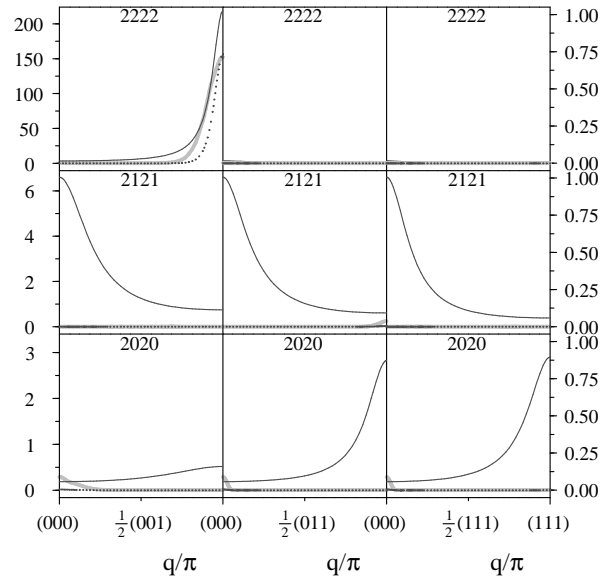}
\caption{\label{fig7}Same as Fig.~\ref{fig3}, but for \textit{prolate} 
  ellipsoids with $a = 1.4524$ and $b = 0.72$, i.e. 
  $(X_{0},\varphi) \protect\cong (2.02,0.394)$.}
\end{figure}
\begin{figure}[ht]
\includegraphics[width=7.5cm,angle =270]{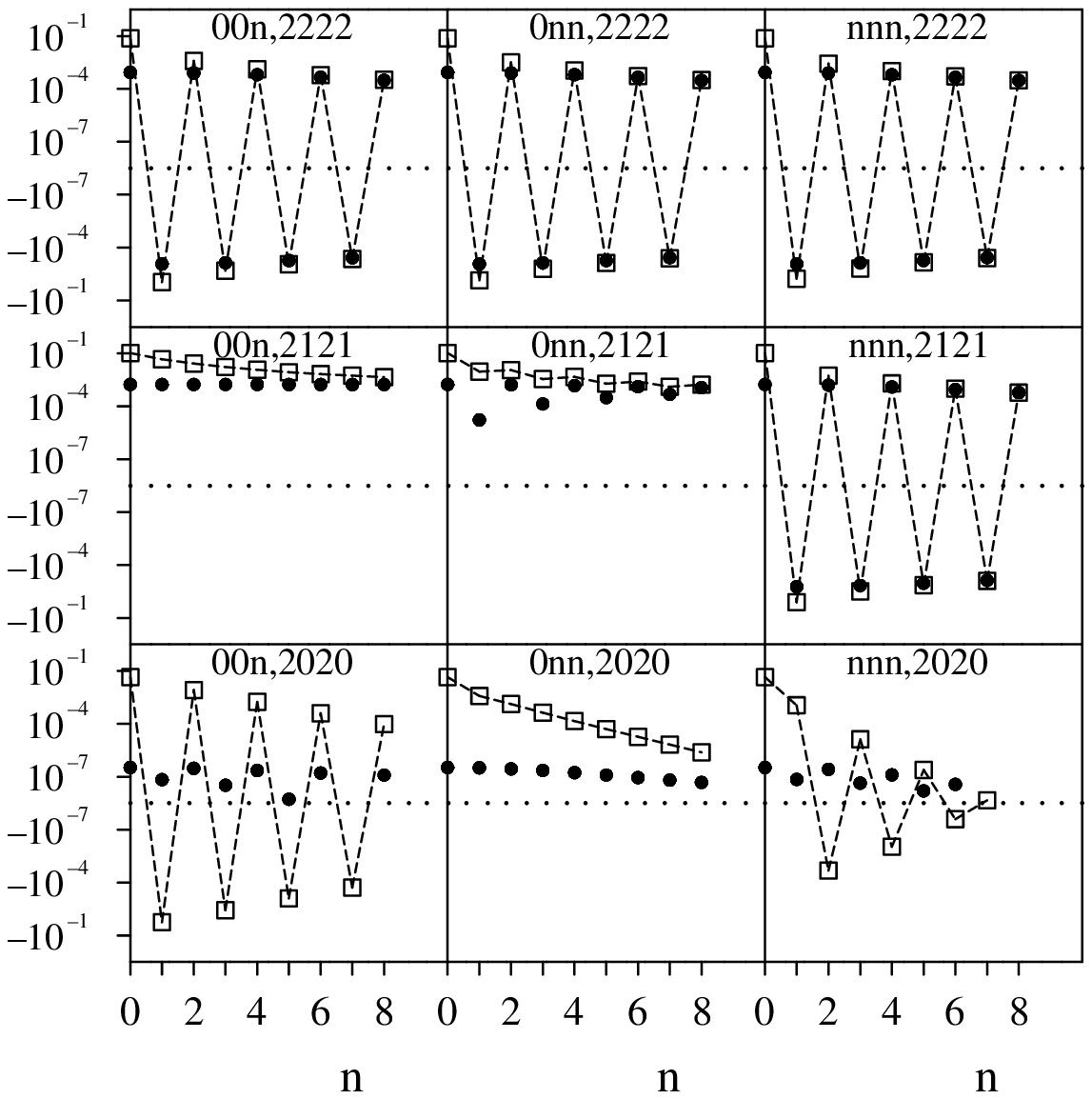}
\caption{\label{fig8}Same as Fig.~\ref{fig4}, but for \textit{prolate} 
  ellipsoids with $a = 1.4524$ and $b = 0.72$, i.e.
  $(X_{0},\varphi) \protect\cong (2.02,0.394)$.}
\end{figure}
\begin{figure}[ht]
\includegraphics[width=11.7cm,angle =270]{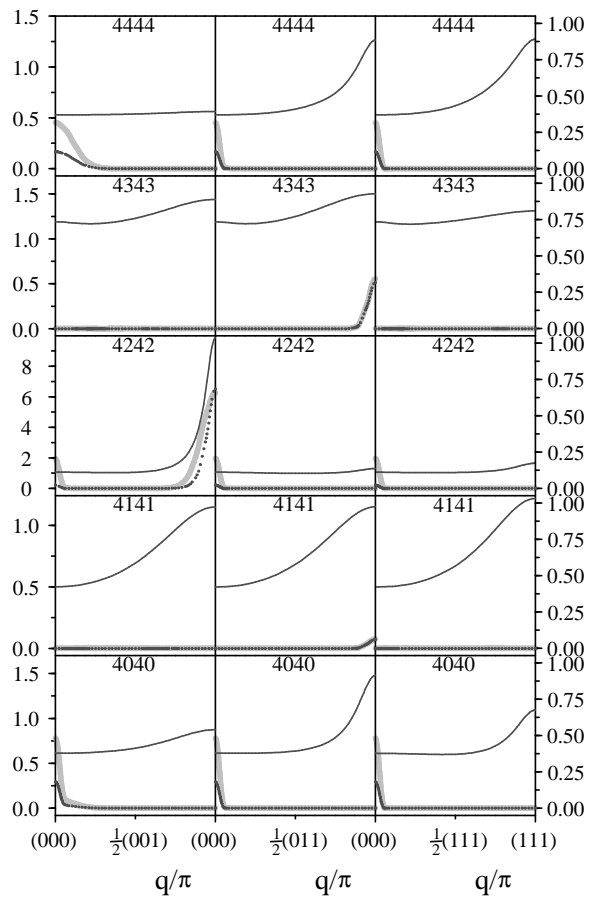}
\caption{\label{fig9}Same as Fig.~\ref{fig3}, but for \textit{prolate} 
  ellipsoids with $a = 1.4524$ and $b = 0.72$, i.e. $(X_{0},\varphi) 
  \protect\cong (2.02,0.394)$, and for $l =l'=4$, $m=m'=0,1,2,3,4$.}
\end{figure}

The $q$-dependence of the critical NEP for \textit{prolate} ellipsoids
is sensitive on the shape of the ellipsoids (see Figs. \ref{fig7},
\ref{fig9} and \ref{fig10}). The reader should note the higher values
of the maxima in $S_{\lambda\lambda'}({\bf q})$, which are necessary
to get a glass transition, compared to the corresponding correlators
for oblate ellipsoids. Let us have a closer look at prolate ellipsoids
with $a = 1.4524$ and $b = 0.72$, i.e. $(X_{0},\varphi) \protect\cong
(2.02,0.394)$. Fig.~\ref{fig7} shows that the structural arrest of
these ellipsoids is completely different from that of oblate ones. The
huge peak in $F^{c}_{22,22}({\bf q}=(0,0,\pi))$ at the zone boundary has 
a height of 157 and dominates the transition. Since this peak belongs 
to a wavelength of period two, it leads to strong frozen oscillations 
in the orientational density fluctuations on the lattice, as can be 
seen in direct space from Fig.~\ref{fig8}. Note that for the correlators
$F^{c}_{00n,21,21}$ almost no decay exists if $n$ is increased. 
Again, like for oblate ellipsoids, the frozen $F^{c}_{xyz,20,20}$ seem 
to play a special role, since they are much weaker than the NEP 
for $m=m'>0$. Fig.~\ref{fig9} shows the diagonal correlators for
$l=l'=4$. Note the very small scale for the static structure factors 
and NEP, in comparison with Fig.~\ref{fig7}. Fig.~\ref{fig9} shows
other interesting features of the MCT results for molecular crystals: 
besides the appearance of
simultaneous maxima of the normalized NEP at the zone center 
and its boundary (see $f_{42,42}({\bf q})$ along the fourfold 
reciprocal space direction), the rule that the normalized NEP 
in reciprocal space are in phase with the corresponding static 
correlators \cite{GS1992} is violated. 

Finally, it must be said that the static structure factors 
for ellipsoids with $a = 1.4524$, $b = 0.72$ have been calculated 
by OZ/PY theory. But MC results \cite{MRic04} for other values of
$(a,b)$ in the vicinity of these parameters show that OZ/PY overestimates 
the maxima at the zone bondary in this region of the phase diagram. 
Therefore, the interpretation of Figs.~\ref{fig7}-\ref{fig9} should be
taken with some caution. Perhaps this overestimation is the indirect cause 
for the dip in $\varphi_{\text{PY}}(X_{0})$ for $2 \lesssim X \lesssim 
4$ (see Fig.~\ref{fig2}). Why OZ/PY fails in this region of 
ellipsoids is currently unknown.

\begin{figure}[ht]
\includegraphics[width=7.5cm,angle =270]{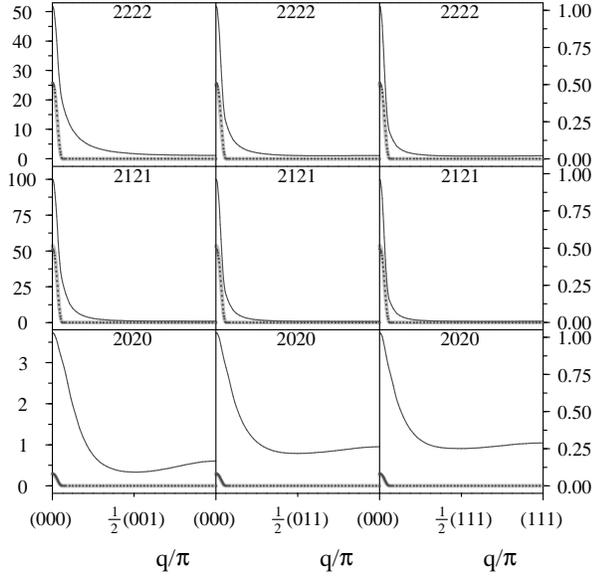}
\caption{\label{fig10}Same as Fig.~\ref{fig3}, but for \textit{prolate} 
  ellipsoids with $a = 5.857$ and $b = 0.24$, i.e.
  $(X_{0},\varphi) \protect\cong (24.4,0.177)$.}
\end{figure}
As we turn to very elongated prolate ellipsoids, the
transition scenario becomes simpler again. Fig.~\ref{fig10} 
for $a = 5.857$ and $b = 0.24$ [yielding $(X_{0},\varphi) \protect\cong 
(24.4,0.177)$] serve as an illustration. The behavior of the $(l=l'=2)$-NEP 
with peaks at the zone center reminds one of the NEP for flat 
oblate ellipsoids (see Fig.~\ref{fig3}). This means that for long 
prolate ellipsoids only nematic-like orientational fluctuations may 
freeze. Such an extreme narrowness of the peaks at ${\bf q} = {\bf
  0}$ as seen in Fig.~\ref{fig10} is observed for prolate ellipsoids
with $X_{0} \gtrsim 8$ only, indicating the huge spatial extension
of the frozen nematic-like fluctuations. 
\subsection{Nonergodicity parameters in the glass phase}
\label{SecIIIC}
In this subsection, we show by means of two examples how the NEP 
change in comparison to the critical NEP on moving slightly into
the glass phase. The corresponding pairs $(X_{0},\varphi)$ are
indicated in Fig.~\ref{fig2}, too. 

\begin{figure}[hb]
\includegraphics[width=7.5cm,angle =270]{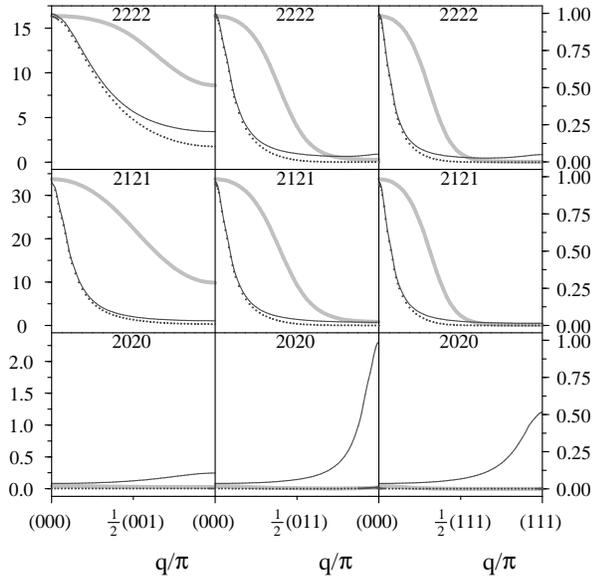}
\caption{\label{fig11}Same as Fig.~\ref{fig3}, but for \textit{oblate} 
  ellipsoids with $a = 0.78$ and $b = 1.1$, i.e.
  $(X_{0},\varphi) \protect\cong (0.709,0.494)$, above the glass line.}
\end{figure}
For densely packed oblate ellipsoids with $a = 0.78$ and $b = 1.1$, 
i.e. $(X_{0},\varphi) \protect\cong (0.709,0.494)$ in the glass phase, 
the prototypical behavior of the critical NEP for oblate ellipsoids
on the glass transition line shown in Figs.~\ref{fig3} and~\ref{fig6} 
is clearly changed, as can be seen from Fig.~\ref{fig11} \footnote{For 
$b=1.1$ we have found the critical $a_{c}\protect\cong 0.7703$. The 
corresponding critical NEP are qualitatively very similar to those
of Fig.~\ref{fig3}. NEP for $a = 1.7$ and $b = 0.66$ are shown in
Fig.~\ref{fig12} and discussed in the next paragraph. For fixed 
$b = 0.66$, $a_{c}\protect\cong 1.6149$. The corresponding critical NEP
are quite similar to those of Fig.~\ref{fig7}.}. Now, the Gaussian-like
shape of the normalized NEP is much broader, indicating an enhanced arrest
of orientational density fluctuations for ${\bf q} \neq {\bf 0}$, which
is expected due to the high packing fraction. This leads to a deviation
of the frozen orientational correlators in direct space from 
the exclusive monotonous decay, which is present almost everywhere
along the glass transition line for oblate ellipsoids. For example, the 
frozen $F_{00n,20,20}$ for the $(a,b)$-pair of Fig. \ref{fig11} (not shown 
here) have weak oscillations, reminiscent of the strong oscillations 
being present in the associated static $G_{00n,20,20}$.

\begin{figure}[h]
\includegraphics[width=7.5cm,angle =270]{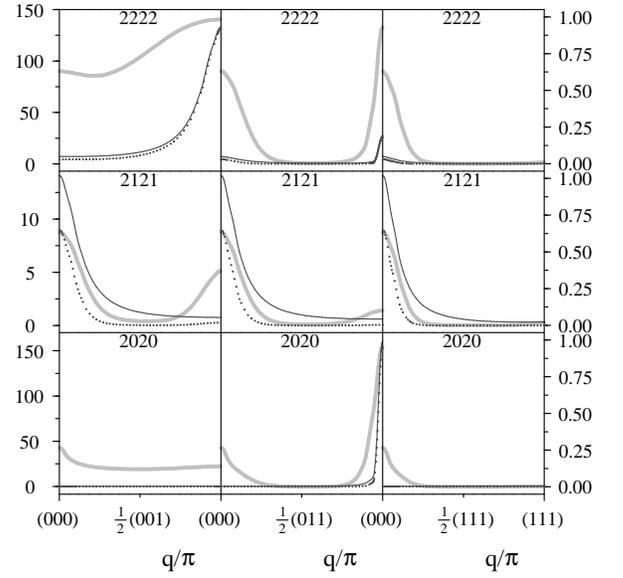}
\caption{\label{fig12}Same as Fig.~\ref{fig3}, but for \textit{prolate} 
  ellipsoids with $a = 1.7$ and $b = 0.66$, i.e.
  $(X_{0},\varphi) \protect\cong (2.58,0.389)$, above the glass line.}
\end{figure}
Considering prolate ellipsoids with $a = 1.7$ and $b = 0.66$, i.e. 
$(X_{0},\varphi) \protect\cong (2.58,0.389)$, slightly above the glass 
transition line, many different patterns of behavior occur in the NEP, 
as can be seen from Fig. \ref{fig12}. This figure can directly be 
compared with Fig.~\ref{fig7}, since the ellipsoids for both figures
have almost the same packing fraction. Again, for one and the 
same NEP there partly exist simultaneous maxima at the zone center
and its boundary. Accordingly, in the limit of long times, we have
frozen density-density correlators with either oscillatory or 
monotonous behavior, depending on $\lambda\lambda'$.
\subsection{Glass transition of the $l,l'$ odd correlators}
\label{SecIIID}
So far we have discussed the NEP for $l,l'$ even. For $l,l'$ odd 
only the ``self'' part of the NEP is nonzero. It is useful to
investigate the normalized, rotationally invariant ``self'' 
part of the NEP, i.e.
\begin{equation}
\label{eq26}
f^{(s)}_{l} = \frac{\sum_{m=-l}^{l} F_{000,lm,lm}}
{\sum_{m=-l}^{l} G_{000,lm,lm}}\,.
\end{equation}
Values for $f^{(s)}_{l}$ are given in Table~\ref{tab1} for those pairs 
$(a,b)$ for which the glass transition has been found for $l,l'$ odd. 
For comparison, the $f^{(s)}_{l}$ for $l$ even are given in 
Table~\ref{tab1}, too. The relation
\begin{equation}
\label{eq27}
f^{(s)}_{l} <  f^{(s)}_{l'}\,, \quad\quad l > l'\,,
\end{equation}
which is similar to
\begin{equation}
\label{eq28}
f^{c}(q) <  f^{c}(q')\,,\quad\quad q > q'
\end{equation}
for simple liquids, seems to be fulfilled for even and odd $l$
separately. Note that the pairs $(a,b)$ in Table \ref{tab1} are
located in the glass phase for $l,l'$ even.
\begin{table}[ht!]
\begin{tabular}{c|cccc}
$(a,b)$ & $l=1$ & $l=2$ & $l=3$ & $l=4$ \\
\hline
$(0.776,1.1)$ & $0.203$  & $0.208$  &  $0.124$ & $0.106$ \\
$(0.778,1.1)$ & $0.406$  & $0.268$  &  $0.262$ & $0.143$ \\
$(0.78,1.1)$ & $0.546$  & $0.333$  &  $0.371$ & $0.185$ \\
$(1.7,0.66)$ & $3.00 \times 10^{-2}$  & $0.197$
& $8.14 \times 10^{-3}$ & $3.93 \times 10^{-2}$ \\
\end{tabular}
\caption{Selected normalized NEP $f^{(s)}_{l}$ of the 
``self'' part of the orientational density-density correlation 
function (see Eqs.~(\ref{eq24}),~(\ref{eq25})).}
\label{tab1}
\end{table}
\section{Discussion and conclusions}
\label{secIV}
We have extended the mode coupling theory (MCT) for liquids to
molecular crystals. The natural choice is the use of tensorial
correlators, instead of correlators defined in a site-site
representation \cite{SHCH98}. This leads for the dynamical
correlators $S_{\lambda \lambda'}({\bf q},t)$ to an
integro-differential equation of second order in time. Truncating
$l$ at $l_\text{max}$, this set of equations is equivalent to the
corresponding equation for a multi-component liquid of isotropic
particles (for binary systems see, e.g., \cite{VG}). 
The memory kernel is approximated in the framework of
MCT. The main differences to liquids are (i) the occurrence of
Umklapp processes, if the sum of the wave vectors ${\bf q}_1$, ${\bf
q}_2$ of the orientational density modes $\delta\rho_{\lambda_1}
({\bf q}_{1})$ and $\delta\rho_{\lambda_2} ({\bf q}_{2})$ is outside 
of the first Brillouin zone, (ii) besides the static two-point
orientational correlators $S_{\lambda \lambda'}({\bf q})$ the
need of the one-molecule orientational density $\rho^{(1)}(\Omega)$
as an input for the vertices of the memory kernel and (iii) the
anisotropy of the static orientational current density correlators 
$J_{\lambda\lambda'}$ which do not cancel completely from the
memory kernel $m_{\lambda \lambda'} ({\bf q}, t)$. Nevertheless,
the factor $k_BT/I$ of $J_{\lambda \lambda'}$ drops out.
Accordingly, the glassy dynamics and the ideal glass transition
does not exhibit inertia effects, i.e. they are independent on
$I$, the moment of inertia. Additionally, for rigid lattices,
all $l,l'=0$ tensorial correlators vanish and can be skipped,
due to the lack of TDOF.

In order to discuss this set of MCT-equations for molecular
crystals we have chosen hard ellipsoids of revolution with aspect
ratio $X_0=a/b$ fixed with their centers of mass at the sites of a
sc-lattice with lattice constant equal to one. Increasing the size
of the ellipsoids, which is equivalent to a decrease of the lattice
constant, results in an increase of steric hindrance and
finally in an orientational glass transition at the MCT-glass
transition line $\varphi_c(X_{0})$ shown in Fig.~\ref{fig2}
for oblate and prolate ellipsoids. Since this orientational glass 
transition is mainly driven by the growth of $S_{\lambda \lambda'} 
({\bf q})$ at the zone center or/and the zone edge, its origin lies 
in the growth of the orientational order close to but below the 
equilibrium phase transition line from OZ/PY theory. This is quite 
similar to what has been found for a liquid of hard ellipsoids 
\cite{MLeRS00} if the aspect ratio becomes larger than about 2 or 
smaller than about $\frac{1}{2}$. However, there is a difference between the 
molecular liquid (of ellipsoids) and the molecular crystal. Whereas
the former already undergoes a glass transition for $X_0 > 2$ or 
$X_0<\frac{1}{2}$ when $S_{2m, 2m} ({\bf 0})$ is of order one, it must 
be $S_{2m,2m} ({\bf 0})$ of order 10 for oblate (cf. Fig.~\ref{fig3})
or even order 100 for prolate ellipsoids (cf. Fig.~\ref{fig7}). This
proves that the translational degrees of freedom of the liquid still
have a strong influence on the glass formation, although they are not
primarily responsible for the transition for $X_{0} > 2$ and $X_{0} <
\frac{1}{2}$. This finding is consistent with results 
found from a MD simulation for difluorotetrachloroethane in its 
supercooled liquid and plastic crystal phase \cite{FAffpri}. For both
phases, the glass transition temperatures $T_{c}^{\text{liquid}}$ and
$T_{c}^{\text{plastic crystal}}$ were determined. That
$T_{c}^{\text{liquid}} \cong 139 \mathrm{K} > T_{c}^{\text{plastic
 crystal}} \cong 129 \mathrm{K}$ implies that the translational
degrees of freedom of a liquid enhance the glass formation which might
be related to a facilitated cage formation for systems where the
center of mass of the particles can move freely.

Comparing $S_{\lambda \lambda'}({\bf q})$ on the glass
transition line for oblate (Figs.~\ref{fig3} 
and~\ref{fig6}) with prolate ellipsoids (Figs.~\ref{fig7} and
\ref{fig9}-\ref{fig10}) already shows that the tendency to an 
orientational glass formation for oblate ellipsoids is larger
than for prolate ones. This can also be seen from Fig.~\ref{fig2}
since the distance $\varphi_{\text{PY}}(X_{0}) - \varphi_{c}(X_{0})$ 
is large for very flat oblate ellipsoids, only. This difference
may be explained as follows. If we fix the length $a$ of prolate
ellipsoids and decrease their thickness $b$ to zero, then the
excluded volume interaction becomes zero. Particularly, the static
correlators become trivial, which leads to vanishing vertices and
consequently to a disappearance of the MCT glass transition. If on
the other hand we fix the diameter $b$ of the oblate ellipsoids
and decrease their thickness $a$ to zero the excluded volume
interaction still exists for $a=0$. This seems to be an important
difference between oblate and prolate particles.

From the solution of the MCT equations for $t \to \infty$ we
obtained results for the critical NEP $F_{\lambda\lambda'}^c 
({\bf q})$ and the corresponding normalized ones, $f_{\lambda
\lambda'}^c ({\bf q})$, as well as NEP deeper in the glass. 
Due to the lattice translational invariance, ${\bf q}$ can
be restricted to the first Brillouin zone. Within this zone 
the critical NEP do not have much structure. Almost all of 
them either exhibit a maximum (or minimum for $l\neq l'$) at 
the zone center and/or at its edge, depending on $\lambda\lambda'$ 
and the direction of ${\bf q}$. However, going deeper into the 
glass and varying the ellipsoid shape, i.e.~$X_{0}$ and/or $\varphi$, 
leads to significant changes in the ${\bf q}$-dependence 
especially of the normalized NEP, as demonstrated in 
Figs.~\ref{fig11} and~\ref{fig12}.

The MD-simulations for cyanoadamantane \cite{AffDes99} and
chloradamantane \cite{AffCoDe01} reveal quite similar glassy
dynamics as found for supercooled liquids
\cite{WGoe98,WGoe99,KBin03}. Particularly, the authors of Refs.
\cite{AffDes99,AffCoDe01} stress that their molecular crystals can
be supercooled and that the relaxational dynamics is consistent
with MCT predictions, at least where this has been checked
\cite{AffDes99,AffCoDe01}. Since both model systems exhibit
tremendous slowing down in the supercooled regime an orientational
glass transition or at least a crossover from ergodic to
quasi-nonergodic dynamics must also occur in the supercooled
phase. This is different from what we have found for the hard
ellipsoids. In our case the glass transition line is located within
the dynamically disordered equilibrium phase of PY
theory, which itself is a consequence of the static input taken
from PY theory. It would be very interesting to insert
the static correlators~\footnote{The simulations in Refs.
\cite{AffDes99,AffCoDe01} are performed for a non-rigid lattice,
i.e. they include phonons. Therefore, ${\bf q}$ is not restricted
to the first Brillouin zone and the ${\bf q}$-dependence of the
static orientational correlators contain a Debye-Waller factor. As
an input into our MCT equations this factor has to be
eliminated.} from the MD simulations in the
supercooled phase into our MCT equations and to check whether one
obtains a glass transition. The investigation of hard
ellipsoids on the sc lattice has demonstrated that the magnitude of
the extrema in $S_{\lambda\lambda'}({\bf q})$ at the zone center
or/and edge must be rather large (at least for prolate particles).
It is not obvious that the simulational results in the supercooled
phase fulfill this criterion. Of course, it could be that the
\textit{average} height of $S_{\lambda \lambda'} ({\bf q})$ in $q$-space
is much larger than for the ellipsoids such that large maxima/minima of
$S_{\lambda\lambda'}({\bm q})$ at the zone center and/or edge are not
really necessary. We hope that these questions can be answered in future.

To conclude, we have shown that MCT can be derived for molecular
crystals, as well. Whether or not the MCT approximation (which
mainly consists of the factorization of a time dependent four
point correlator) is also a reasonable approximation for molecular
crystals as it is for glass-forming liquids has to be investigated
by comparison of the results from present MCT for molecular
crystals with simulational and experimental results. As already
mentioned above our conventional MCT approach will become worse if
the thickness of \textit{prolate} particles becomes small. In that
case it is the entanglement which is responsible for glassy
dynamics \cite{CRenLoe97,SPOb97}. This requires a different
theoretical description, as recently discussed
\cite{RSGZ03,RSAd04}.
\appendix
\section{Calculation of $J_{\lambda \lambda'} ({\bf q})\equiv J_{\lambda
\lambda'}$}
\label{appA}
Substituting the $\lambda$-Fourier transform $j_\lambda ({\bf q})$
of $j_n(\Omega)$ (Eq.~(\ref{eq6})) into Eq.~(\ref{eq11}) yields
\begin{align}
\label{eqA1} 
&J_{\lambda\lambda'}({\bf q}) = \frac{4\pi}{N}\,i^{l'-l}\,
\sum_{nn'} \, \text{e}^{i{\bf q} \cdot{\bf x}_{nn'}} \,\times
\nonumber\\
&\times\,\big\langle \, ({\bm \omega}_{n} \cdot
\hat{{\bf L}}_{\Omega_{n}} Y_{\lambda}(\Omega_{n}))^{*} 
\,({\bm \omega}_{n'} \cdot  \hat{{\bf L}}_{\Omega_{n'}}
Y_{\lambda'}(\Omega_{n'})) \,\big\rangle\,.
\end{align}
Since $\boldsymbol{\omega}_n \cdot \hat{{\bf L}}_{\Omega_n}= 
\boldsymbol{\omega}'_n \cdot \hat{{\bf L}}'_{\Omega_n}$ (where 
the primed quantities refer to the body fixed frame) we get for 
the canonical average in Eq.~(\ref{eqA1}) in close analogy 
to molecular liquids \cite{RSTS97}
\begin{align}
\label{eqA2}
\big\langle &\ldots\big\rangle = \sum_{\alpha\alpha'}\,
\big\langle \omega_{n}'^{\alpha} \omega_{n'}'^{\alpha'} \big\rangle \,
\big\langle \,(\hat{L'}_{\Omega_{n}}^{\alpha} \,Y_{\lambda}(\Omega_{n}))^{*}\,
(\hat{L'}_{\Omega_{n'}}^{\alpha'} \,Y_{\lambda'}(\Omega_{n'}))\,
\big\rangle \,\nonumber\\
& = \sum_{\alpha\alpha'}\, \frac{kT}{I} \, \delta_{nn'}
\delta_{\alpha\alpha'} \, \big\langle \,(\hat{L'}_{\Omega_{n}}^{\alpha}
Y_{\lambda}(\Omega_{n}))^{*} \,(\hat{L'}_{\Omega_{n'}}^{\alpha'}
Y_{\lambda'}(\Omega_{n'}))\,\big\rangle\, ,
\end{align}
where, $\alpha = x,y,z$ are the cartesian components 
in the body fixed frame. This leads to
\begin{align}
\label{eqA3}
 J_{\lambda\lambda'}({\bf q}) & = 
4\pi\,\frac{kT}{I}\,i^{l'-l}\,\big\langle \,(\hat{{\bf L}}_{\Omega_{n}}
Y_{\lambda}(\Omega_{n}))^{*} \, \cdot\,(\hat{{\bf L}}_{\Omega_{n}}
Y_{\lambda'}(\Omega_{n}))\,\big\rangle \,,
\end{align}
which is ${\bf q}$- and $n$-independent, and can therefore
be evaluated for arbitrary $n$.
\begin{widetext}
Using $\hat{L}^\alpha_\Omega \,Y_{lm}(\Omega)= \sum_{m'=-l}^{l} 
L^\alpha_{l, mm'} Y_{l m'}(\Omega)$ and the product rule for the 
spherical harmonics and substituting the explicit expression 
for $L^\alpha_{l, m m'}$ we get with $c(l,l',l'')$ as in~(\ref{eqC4})
\begin{align}
\label{eqA4}
J_{\lambda\lambda'}({\bf q}) =\:&4\pi\,\frac{kT}{I}\, i^{l'-l} \, 
(-1)^{m}\, \sum_{\lambda''} \Big[ \, mm' \,C(ll'l'',-mm'm'')
\nonumber\\
&-\frac{1}{2} \sqrt{l(l+1)-m(m+1)}\,\sqrt{l'(l'+1)-m'(m'+1)}\;
C(ll'l'',-m-1,m'+1,m'')
\nonumber\\[1ex]
&-\frac{1}{2} \sqrt{l(l+1)-m(m-1)}\,\sqrt{l'(l'+1)-m'(m'-1)}\;
C(ll'l'',-m+1,m'-1,m'') \,\Big] \, c(ll'l'') \, \big\langle \, 
Y_{\lambda''} \, \big\rangle\,.
\end{align}
This expression strongly simplifies since
\begin{align}
\label{eqA5}
[ \cdots ] \,c(ll'l'') & = -\frac{1}{2}\,
\sqrt{l(l+1)}\,\sqrt{l'(l'+1)}\,\left[
\frac{(2l+1)(2l'+1)}{4\pi\,(2l''+1)} \right]^{\frac{1}{2}}
\big\{ C(ll'l'',1-10)+C(ll'l'',-110)\big\} \,
C(ll'l'',-mm'm'') \,.
\end{align}
This leads to the final result
\begin{align}
\label{eqA6}
J_{\lambda\lambda'}({\bf q})&=
4\pi\,\frac{kT}{I}\, i^{l'-l}\, (-1)^{m+1} \, \frac{1}{2}\,
\sqrt{l(l+1)}\,\sqrt{l'(l'+1)}\,\times
\nonumber\\[1ex]
&\times \sum_{\lambda''} \left[
\frac{(2l+1)(2l'+1)}{4\pi\,(2l''+1)} \right]^{\frac{1}{2}}
\big\{ C(ll'l'',1-10)+C(ll'l'',-110)\big\}
\, C(ll'l'',-mm'm'') \, \big\langle \,Y_{\lambda''} \, \big\rangle\,.
\end{align}
\end{widetext}
Note that $\langle Y_\lambda\rangle$ is given by
\begin{equation}
\big\langle \, Y_{\lambda} \, \big\rangle = \int\, \text{d} 
\Omega\,\rho^{(1)}(\Omega) \,Y_{\lambda}(\Omega)  = (-i)^{l}\,
\rho^{(1)}_{\lambda}\,,
\end{equation}
i.e.~$J_{\lambda \lambda'} ({\bf q}) \equiv J_{\lambda \lambda'}$
only involves the $\lambda$-transform of $\rho^{(1)}(\Omega)$.
\section{Mode coupling approximation}
\label{appB}
In this appendix we shortly describe the mode coupling
approximation leading to the results presented by
Eqs.~(\ref{eq15})-(\ref{eq19}).

The derivation of the Mori-Zwanzig equation is standard by using
the projectors onto the slow variables $\delta \rho_\lambda({\bf
q})$ and $j_\lambda ({\bf q})$,
\begin{align}
\label{eqB1}
P_{\rho} &= \frac{4\pi}{N} \,\sum_{\lambda\lambda'}\!'\, \,
\big| \,\delta \rho_{\lambda}({\bf q}) \,\big\rangle \,
({\bf S}^{-1}({\bf q}))_{\lambda\lambda'}
\,\big\langle \, \delta \rho^{*}_{\lambda'}({\bf q})\,\big|\,,\\[1ex]
\label{eqB2} P_{j} &= \frac{4\pi}{N}
\,\sum_{\lambda\lambda'}\!'\,  \, \big| \,j_{\lambda}({\bf q})
\,\big\rangle \,({\bf J}^{-1} )_{\lambda\lambda'} \,\big\langle \,
j^{*}_{\lambda'}({\bf q})\,\big|\,.
\end{align}
The prime on sums denotes summation such that $l,l'>0$.
The projector $Q$ in Eq.~(\ref{eq14}) is the given by
\begin{equation}
\label{eqB3} Q = 1 - P_{\rho} - P_{j}\,.
\end{equation}
In order to approximate $M_{\lambda \lambda'} ({\bf q}, t)$
Eq.~(\ref{eq14}) we introduce the projector onto pairs of
orientational density modes:
\begin{align}
\label{eqB4}
&\mathcal{P} = \sum_{\substack{{\bf q}_{1} {\bf q}_{1}',
  {\bf q}_{2} {\bf q}_{2}' \\ \in \,\text{1.BZ}}}
\;\sum_{\lambda_{1}\lambda_{1}'\lambda_{2}\lambda_{2}'}\hspace{-2ex}'\;
\;\big| \, \delta\rho_{\lambda_{1}}({\bf q}_{1}) \,
\delta\rho_{\lambda_{1}'}({\bf q}_{1}')
\, \big\rangle \,\times
\nonumber\\[1.5ex]
&\times \, g_{\lambda_{1}\lambda_{1}';\lambda_{2}\lambda_{2}'}
({\bf q}_{1}{\bf q}_{1}',{\bf q}_{2}{\bf q}_{2}') \;
\big\langle \, \delta\rho^{*}_{\lambda_{2}}({\bf q}_{2}) \, 
\delta\rho^{*}_{\lambda_{2}'}({\bf q}_{2}') \, \big|\,,
\end{align}
where $(g_{\lambda_1 \lambda'_1, \lambda_2 \lambda'_2} ({\bf q}_1
{\bf q}'_1; {\bf q}_2 {\bf q}'_2))$ is the inverse of the static
four-point correlation matrix $(\langle \,\delta\rho^*_{\lambda_1}
({\bf q}_1) \,\delta \rho^*_{\lambda'_1} ({\bf q}'_1) \,\times$ 
$\times\,\delta \rho_{\lambda_2} ({\bf q}_2) \,\delta\rho_
{\lambda'_2}({\bf q}'_2 )\,\rangle)$. We use the approximation
\begin{align}
\label{eqB5}
&g_{\lambda_{1}\lambda_{1}';\lambda_{2}\lambda_{2}'}({\bf q}_{1}{\bf
 q}_{1}',{\bf q}_{2}{\bf q}_{2}') \approx \frac{1}{4} \,
\left(\frac{4 \pi}{N} \right)^{2}\times\\[1.5ex]
&\times\Big[ \,\delta_{{\bf
q}_{1}{\bf q}_{2}} \,\delta_{{\bf q}_{1}',{\bf q}_{2}'} \,
({\bf S}^{-1}({\bf q}_{1}))_{\lambda_{1}\lambda_{2}} \,
({\bf S}^{-1}({\bf q}_{2}))_{\lambda'_{1}\lambda'_{2}}
+ (1 \leftrightarrow 2) \,\Big] ,\nonumber
\end{align}
which is consistent with the mode coupling approximation of
$M_{\lambda \lambda'} ({\bf q}, t)$ for $t=0$ (see Eq.~(\ref{eqB7})).

The mode coupling approximation consists of two main steps. First,
the fluctuating force is approximated
\begin{equation}
\label{eqB6} Q\;\big|\,\mathcal{L} j_{\lambda}({\bf q}) \,
\big\rangle \approx \mathcal{P} Q \;\big|\,\mathcal{L} 
j_{\lambda}({\bf q}) \big\rangle \,.
\end{equation}
Substituting (\ref{eqB6}) into Eq.~(\ref{eq14}) leads to a time-dependent
four-point correlator, which in a second approximation is
factorized. For ${\bf q}_{1},{\bf q}'_{1},{\bf q}_{2},{\bf
  q}'_{2} \in$ 1.BZ we have
\begin{align}
\label{eqB7}
&\big\langle \,\delta\rho^{*}_{\lambda_{1}}({\bf q}_{1})
\,\delta \rho^{*}_{\lambda_{1}'}({\bf q}_{1}') \,\big|
 \,Q\,\text{e}^{-i \,  Q\,\mathcal{L} \, Q \, t} \, Q \,
\, \big| \,\delta\rho_{\lambda_{2}}({\bf q}_{2})
\,\delta\rho_{\lambda_{2}'} ({\bf q}_{2}') \,\big\rangle
\approx\nonumber\\[1.5ex]
& \frac{N^{2}}{(4\pi)^{2}} \, \Big[ \, \,\delta_{{\bf
q}_{1},{\bf q}_{2}} \,\delta_{{\bf q}_{1}',{\bf q}_{2}'}
\,S_{\lambda_{1}\lambda_{2}}({\bf q}_{1},t)
\,S_{\lambda_{1}'\lambda_{2}'}({\bf q}_{2},t)+ 
(1 \leftrightarrow 2) \, \Big].
\end{align}

With these approximations we obtain
\begin{align}
\label{eqB8} 
M_{\lambda\lambda'}&({\bf q},t) \approx 
\frac{1}{2}\,\left(\frac{4\pi}{N}\right)^{3} \, \sum\limits_{{\bf
q}_{1}{\bf q}_{2} \atop \in \text{1.BZ}} \; \sum\limits_{\substack{
\lambda_{1}\lambda_{2}\lambda_{3}\lambda_{4} \\
\lambda_{1}'\lambda_{2}'\lambda_{3}'\lambda_{4}'}}
\hspace{-2.5ex}'  \nonumber\\
&\times\big\langle\, (\mathcal{L}
\, j_{\lambda} ({\bf q}) )^{*} \,\big| \, Q \, \big| \,
\delta\rho_{\lambda_{1}}({\bf q}_{1}) \, \delta\rho_{\lambda_{2}}
({\bf q}_{2}) \,\big\rangle \, \times
\nonumber\\[1.5ex]
&\times ( {\bf S}^{-1} ({\bf q}_{1})
)_{\lambda_{1}\lambda_{3}} \, ( {\bf S}^{-1} ({\bf
q}_{2}) )_{\lambda_{2}\lambda_{4}} \times
\nonumber\\[1.5ex]
&\times S_{\lambda_3 \lambda'_3} ({\bf q}_1, t) \,S_{\lambda_4 \lambda'_4}
 ({\bf q}_2, t)\,\times
\nonumber\\[1.5ex]
&\times ( {\bf S}^{-1}
({\bf q}_{1}) )_{\lambda_{3}'\lambda_{1}'} \, ( {\bf
S}^{-1} ({\bf q}_{2}))_{\lambda_{4}'\lambda_{2}'} \times
\nonumber\\[1.5ex]
&\times\big\langle \,\delta\rho^{*}_{\lambda_{2}'}({\bf q}_{2})
 \,\delta \rho^{*}_{\lambda_{1}'}({\bf q}_{1})\,\big| \,Q\,\big| 
\,\mathcal{L} \,j_{\lambda'}({\bf q}) \,\big\rangle \,.
\end{align}
\section{Calculation of $\langle \,
  ( \mathcal{L} j_{\lambda} ({\bf q}) )^{*} \, | \, Q \, | \,
  \delta\rho_{\lambda_{1}} ({\bf q}_{1})\,
\delta\rho_{\lambda_{2}} ({\bf q}_{2}) \, \rangle$}
\label{appC}
This correlator is calculated quite similar to simple
\cite{WGoe91} and molecular liquids \cite{RSTS97,RSP02,LFa99} by
using Eq.~(\ref{eqB3}) and $P_j\,|\,\delta \rho_{\lambda_1} ({\bf
q}_1) \,\delta \rho_{\lambda_2} ({\bf q}_2) \,\rangle=0$, due to time
reversal symmetry. Then we get
\begin{align}
\label{eqC1}
&\big\langle \,
  ( \mathcal{L} j_{\lambda} ({\bf q}) )^{*} \, \big| \, Q \, 
\big| \, \delta\rho_{\lambda_{1}} ({\bf q}_{1})\,
\delta\rho_{\lambda_{2}} ({\bf q}_{2}) \, \big\rangle =
\nonumber \\[1ex]
&=\big\langle \, ( \mathcal{L} j_{\lambda} ({\bf q}) )^{*} \,
  \delta\rho_{\lambda_{1}} ({\bf q}_{1})\,
\delta\rho_{\lambda_{2}} ({\bf q}_{2}) \, \big\rangle 
\nonumber \\[1ex]
&-\big\langle \, ( \mathcal{L} j_{\lambda} ({\bf q}) )^{*} \, \big| \,
P_{\rho} \, \big| \, \delta\rho_{\lambda_{1}} ({\bf q}_{1})\,
\delta\rho_{\lambda_{2}} ({\bf q}_{2}) \, \big\rangle \,.
\end{align}
The \textit{first} term on the r.h.s. of Eq.~(\ref{eqC1}) is easily
rewritten by taking into account the hermiticity of ${\cal L}$ and
the continuity equation Eq.~(\ref{eq4}) and Eq.~(\ref{eq6}). This
leads to
\begin{align}
\label{eqC2}
&\big\langle \,
  ( \mathcal{L} j_{\lambda} ({\bf q}) )^{*} \,
  \delta\rho_{\lambda_{1}} ({\bf q}_{1})\,
\delta\rho_{\lambda_{2}} ({\bf q}_{2}) \, \big\rangle = 
\nonumber \\[1ex]
&=\big\langle \,j^{*}_{\lambda} ({\bf q}) \,
  j_{\lambda_{1}} ({\bf q}_{1})\,
\delta\rho_{\lambda_{2}} ({\bf q}_{2}) \, \big\rangle + ( 1
\leftrightarrow 2) \,.
\end{align}
\begin{widetext}
Substituing the $\lambda$-Fourier transform of $\delta
\rho_n(\Omega)$ and $j_n (\Omega)$ into the r.h.s of
Eq.~(\ref{eqC2}) we arrive at
\begin{align}
\label{eqC3}
\big\langle \, j^{*}_{\lambda} ({\bf q}) \,
j_{\lambda_{1}} ({\bf q}_{1})  \,\delta\rho_{\lambda_{2}} ({\bf
  q}_{2}) \, \big\rangle = &\frac{N}{4\pi}\,\frac{kT}{I}\, 
\sum_{{\bf Q}} \delta_{{\bf q}_{1} + {\bf q}_{2},{\bf q}+{\bf Q}}
\sum_{\lambda''}\!'\;i^{l_{1}+l''-l} \,\,(-1)^{m+m''}\, 
\bigg[ \,mm_{1} \,  C(ll_{1}l'',-mm_{1}-m'')
\nonumber\\[-0.2ex]
-\frac{1}{2} \sqrt{l(l+1)-m(m+1)}\, &\sqrt{l_{1}(l_{1}+1)-m_{1}(m_{1}+1)}\;
C(ll_{1}l'',-m-1,m_{1}+1,-m'')
\nonumber\\[0.8ex]
-\frac{1}{2} \sqrt{l(l+1)-m(m-1)}\, &\sqrt{l_{1}(l_{1}+1)-m_{1}(m_{1}-1)}\;
C(ll_{1}l'',-m+1,m_{1}-1,-m'')\,\bigg] \,
c(ll_{1}l'') \, S_{\lambda''\lambda_{2}}({\bf q}_{2})
\end{align}
with
\begin{equation}
\label{eqC4} c(ll'l'') = \left[ \frac{(2l+1)(2l'+1)}{4\pi\,
(2l''+1)} \right]^{\frac{1}{2}} C(ll'l'',000) \, .
\end{equation}
Here we have used the product rule for the spherical harmonics and
the factorization of canonic integrals as in Eq. (\ref{eqA2}). 
The \textit{second} term on the r.h.s. of
Eq.~(\ref{eqC1}) is rewritten by using $P_\rho$ from Eq.~(\ref{eqB1}) 
and again the hermiticity of ${\cal L}$, as well as the 
continuity equation:
\begin{align}
\label{eqC5}
\big\langle \, ( \mathcal{L} & j_{\lambda} ({\bf q}) )^{*} \, \big|
\, P_{\rho} \, \big| \, \delta\rho_{\lambda_{1}} ({\bf q}_{1})\,
\delta\rho_{\lambda_{2}} ({\bf q}_{2}) \, \big\rangle =
\sum_{\lambda'\lambda''}\!' \;
J_{\lambda\lambda'} \, ( {\bf S}^{-1} ({\bf q})
)_{\lambda'\lambda''} \, \big\langle \,\delta
\rho^{*}_{\lambda''}({\bf q}) \, \delta\rho_{\lambda_{1}} ({\bf
q}_{1})\, \delta\rho_{\lambda_{2}} ({\bf q}_{2}) \, \big\rangle \, .
\end{align}
Substituting Eqs.~(\ref{eqC2}),~(\ref{eqC3}) and~(\ref{eqC5})
into Eq.~(\ref{eqC1}), the l.h.s of Eq.~(\ref{eqC1}) is expressed 
by the static two-point and three-point correlators $S_{\lambda
\lambda'} ({\bf q})$ and $\langle \,\delta \rho^{*}_{\lambda ''} ({\bf
q}) \,\delta \rho_{\lambda_1} ({\bf q}_1) \,\delta
\rho_{\lambda_2} ({\bf q}_2) \,\rangle$, respectively, and by
$J_{\lambda \lambda'}.$ ${\bf J}=(J_{\lambda \lambda'})$ is
calculated in App. \ref{appA}, $\langle \,\delta \rho^{*}_{\lambda ''} 
({\bf q}) \,\delta \rho_{\lambda_1} ({\bf q}_1)\, \delta
\rho_{\lambda_2} ({\bf q}_2) \,\rangle$ in App.~\ref{appD}.

Now we rewrite $\sum_{\lambda''}\!' \ldots$ in Eq.~(\ref{eqC3}) as follows:
\begin{align}
\label{eqC6}
\sum_{\lambda''}\!' \ldots \,=\,
\sum_{\lambda''}\!'\,\sum_{\lambda_{1}'}\hspace{0ex}'\,i^{l_{1}+l''-l} \, &
\,(-1)^{m+m''}\, \bigg[ \,mm'_{1} \,  C(ll'_{1}l'',-mm'_{1}-m'')
\nonumber\\[-0.5ex]
-\frac{1}{2} \sqrt{l(l+1)-m(m+1)}\, &
\sqrt{l'_{1}(l'_{1}+1)-m'_{1}(m'_{1}+1)}\;
C(ll'_{1}l'',-m-1,m'_{1}+1,-m'')
\nonumber\\[0.5ex]
-\frac{1}{2} \sqrt{l(l+1)-m(m-1)}\, &
\sqrt{l'_{1}(l'_{1}+1)-m'_{1}(m'_{1}-1)}\;
C(ll'_{1}l'',-m+1,m'_{1}-1,-m'')\,\bigg] \, \times
\nonumber\\[0.5ex]
\times\, c(ll'_{1}l'') \,& \sum_{\lambda'''}
\,({\bf S}^{-1} ({\bf q}_{1}) )_{\lambda_{1}'\lambda'''}
\,S_{\lambda'''\lambda_{1}}({\bf q}_{1})
\,S_{\lambda''\lambda_{2}}({\bf q}_{2})\,,
\end{align}
and substitute succesively the terms on the r.h.s. of
\begin{align} 
\label{eqC7}
{\bf S}^{-1}({\bf q}_{1}) = {\bf d}^{-1} - {\bf d}^{-1} +  
{\bf D}^{-1}  - \frac{1}{4\pi} \, {\bf c}({\bf q}_{1})\,,
\end{align}
which is a rearrangement of the OZ equation~(\ref{eq20}), 
into Eq.~(\ref{eqC6}).

In the first step we replace $d^{-1}_{\lambda_{1}'\lambda'''} \,
S_{\lambda'''\lambda_{1}}({\bf q}_{1}) \, S_{\lambda''\lambda_{2}}
({\bf q}_{2}) = d_{\lambda''\lambda'_{2}} \, (\,d^{-1}\,S\,)
_{\lambda_{1}'\lambda_{1}} ({\bf q}_{1}) \, (\,d^{-1} S\,
)_{\lambda'_{2}\lambda_{2}}({\bf q}_{2})$. This expression arises
if the matrix elements of ${\bf d}^{-1}$ on the r.h.s. of (\ref{eqC7}) 
are used with (\ref{eqC6}). Using the explicit result for the matrix 
${\bf d}$ (see \cite{MRic04}) and the relations
\begin{align} 
\label{eqC8}
\sum\limits_{\lambda''} \big[ &mm_{1}'\,c(ll_{1}'l'') \,
  C(ll_{1}'l'',-mm_{1}'m'') \, c(l''l_{2}'l') \, C(l''l_{2}'l',m''m_{2}'m')
\nonumber\\
+\,& mm_{2}'\,c(ll_{2}'l'') \, C(ll_{2}'l'',-mm_{2}'m'') \, c(l''l_{1}'l')
\, C(l''l_{1}'l',m''m_{1}'m')\,\big]
\nonumber\\[1.5ex]
= \sum\limits_{\lambda''} &\, mm''c(l_{1}'l_{2}'l'') \, 
C(l_{1}'l_{2}'l'',m_{1}'m_{2}'m'') \, c(ll''l')\, C(ll''l',-mm''m')\,,
\end{align}
\begin{align}
\label{eqC9}
\sum\limits_{\lambda''} \big[ 
&\sqrt{l(l+1)-m(m \mp 1)} \, \sqrt{l_{1}'(l_{1}'+1)-m_{1}'(m_{1}' \mp 1)}
\,c(ll_{1}'l'') \, C(ll_{1}'l'',-m \pm 1,m_{1}' \mp 1,m'') \, c(l''l_{2}'l')
\, C(l''l_{2}'l',m''m_{2}'m')
\nonumber\\
+ \,&\sqrt{l(l+1)-m(m \mp 1)} \, \sqrt{l_{2}'(l_{2}'+1)-m_{2}'(m_{2}' \mp 1)}
\,c(ll_{2}'l'') \, C(ll_{2}'l'',-m \pm 1,m_{2}' \mp 1,m'') \, c(l''l_{1}'l')
\, C(l''l_{1}'l',m''m_{1}'m')\,\big]
\nonumber\\[1.5ex]
= \sum\limits_{\lambda''} &\, \sqrt{l(l+1)-m(m \mp 1)} \, 
\sqrt{l''(l''+1)-m''(m'' \mp 1)} \,c(l_{1}'l_{2}'l'') \, 
C(l_{1}'l_{2}'l'',m_{1}'m_{2}'m'') \, c(ll''l')
\, C(ll''l',-m \pm 1,m'' \mp 1,m')\,,
\end{align}
we find that this part of (\ref{eqC6}) taken together with the same
part in the partner expression of (\ref{eqC6}) due to (\ref{eqC2})
cancels with the part $\langle \, ( \mathcal{L} j_{\lambda} ({\bf q}) 
)^{*} \, | \,P_{\rho} \, | \, \delta\rho_{\lambda_{1}} ({\bf q}_{1})\,
\delta\rho_{\lambda_{2}} ({\bf q}_{2}) \, \rangle$ of (\ref{eqC1}),
if Eqs.~(\ref{eqD8}) and~(\ref{eqA4}) are used in Eq.~(\ref{eqC5}).

We turn to the term ${\bf D}^{-1} -{\bf d}^{-1}$ of
(\ref{eqC7}), which leads to $( {\bf D}^{-1} -{\bf d}^{-1})_
{\lambda_{1}'\lambda'''} S_{\lambda'''\lambda_{1}}({\bf q}_{1})$ 
if substituted in (\ref{eqC6}). Since 
\begin{align}
\label{eqC10} 
\left[D^{-1}-d^{-1}\right](\Omega,\Omega')=
\frac{1}{4\pi} \left[ - \frac{1}{4\pi} \, \frac{1}{\rho^{(1)}(\Omega)}
- \frac{1}{4\pi} \, \frac{1}{\rho^{(1)}(\Omega')} +
\frac{1}{(4\pi)^{2}} \, \int_{S^{2}} \, \frac{1}{\rho^{(1)}(\Omega)} \,
\text{d} \Omega \, \right]\,,
\end{align}
${\bf D}^{-1} -{\bf d}^{-1}$ consists just of a nontrivial first row 
and column, while ${\bf S}({\bf q}_{1})$ has a vanishing first row and
column. So the product $({\bf D}^{-1} -{\bf d}^{-1})\, {\bf S}
({\bf q}_{1})$ has nonvanishing elements in its first row, only. Therefore,
$( {\bf D}^{-1} -{\bf d}^{-1})_{\lambda_{1}'\lambda'''} S_{\lambda'''
\lambda_{1}}({\bf q}_{1}) = 0$, if not $l_{1}' = m_{1}' =0$. But if we
evaluate the coefficients of (\ref{eqC6}) with $l_{1}' = m_{1}' =0$,
we find that the part ${\bf D}^{-1} -{\bf d}^{-1}$ of (\ref{eqC7})
contributes nothing.

What remains is the last term on the r.h.s. of (\ref{eqC7}). If 
substituted into Eq.~(\ref{eqC6}) and the corresponding partner expression
due to Eq.~(\ref{eqC2}), respectively, this term delivers the final result
\begin{align} 
\label{eqC11} 
\big\langle &\, ( \mathcal{L} j_{\lambda}
({\bf q}))^{*} \,\big| \,Q\,\big|\, \delta\rho_{\lambda_{1}} ({\bf
q}_{1})\, \delta\rho_{\lambda_{2}} ({\bf q}_{2}) \, \big\rangle  = 
\nonumber\\[1ex]
&= -\frac{N}{(4\pi)^{\frac{5}{2}}} \, \frac{kT}{I} \,\sum_{{\bf Q}} 
\delta_{{\bf q}_{1} + {\bf q}_{2},{\bf q}+{\bf Q}} 
\sum\limits_{\lambda_{1}'\lambda_{2}'\lambda_{3}'}\hspace{-1ex}'
 \,\left[ \,b(l_{3}'l_{2}'l) \, C(l_{3}'l_{2}'l,m_{3}'m_{2}'m) \,
c_{\lambda_{3}'\lambda_{1}'} ({\bf q}_{1})\,
S_{\lambda_{1}'\lambda_{1}} ({\bf q}_{1})\,
S_{\lambda_{2}'\lambda_{2}} ({\bf q}_{2})
+ (1 \leftrightarrow 2) \,\right],
\end{align}
with $b(l,l',l'')$ from Eq.~(\ref{eq18}). Here we have used the 
relation (\ref{eqA5}) for the Clebsch-Gordan-coefficients. If 
Eq.~(\ref{eqC11}) and its conjugate is substituted into
Eq.~(\ref{eqB8}) one obtains the mode coupling approximation of the 
slow part of $M_{\lambda\lambda'} ({\bf q}, t)$, which then leads to
the final result for ${\bf m}({\bf q}, t)$, Eqs.~(\ref{eq15})-(\ref{eq19}).
\end{widetext}
\section{Approximation of $\langle \,\delta \rho^*_{\lambda}({\bf q}_1)
 \delta \rho_{\lambda_2}({\bf q}_{2})  \delta
\rho_{\lambda_3}({\bf q}_{3}) \,\rangle$}
\label{appD}
The approximation of the static three-point correlator is rather
involved. Therefore, the most crucial steps are presented
only. Readers which are interested in more details are referred to 
Ref. \cite{MRicker04}.

The corresponding static three-point-correlator for \textit{simple}
liquids was approximated by the convolution approximation
\cite{WGoe91}. It has been proven that the
approximation of the corresponding correlator for molecular
liquids \cite{RSTS97} is again the convolution approximation as
defined in Ref. \cite{F1967}. However, performing the convolution
approximation for molecular crystals does not lead to a simple
result. Therefore, we have chosen a different approximation. 
$\langle\,\delta \rho^{*}_{\lambda_1}({\bf q}_1)\,\delta\rho_{\lambda_2}
({\bf q}_2) \,\delta\rho_{\lambda_3} ({\bf q}_3) \,\rangle$ is the 
$\lambda$-Fourier transform of $\langle\,\delta\rho_{n_{1}}(\Omega_{1})\,
\delta\rho_{n_{2}}(\Omega_{2})\, \delta\rho_{n_{3}}(\Omega_{3})\,
\rangle$ given by:
\begin{align}
\label{eqD1}
\big\langle\,\delta&\rho^{*}_{\lambda_1}({\bf q}_1)\,\delta\rho_{\lambda_2}
({\bf q}_2) \,\delta\rho_{\lambda_3} ({\bf q}_3) \,\big\rangle = 
\nonumber\\[3ex]
=&\sum\limits_{n_{1}n_{2}n_{3}} \text{e}^{i(-{\bf q}_{1}
\cdot{\bf x}_{n_1}+ {\bf q}_{2}\cdot{\bf x}_{n_2} +
{\bf q}_{3}\cdot{\bf x}_{n_3})}\,i^{l_{2}+l_{3}-l_{1}} \times
\nonumber\\ 
&\times\iiint \text{d}\Omega_{1}\,\text{d}\Omega_{2} \,\text{d}
\Omega_{3}\,Y_{\lambda_{1}}^{*}(\Omega_{1})\,
Y_{\lambda_{2}}(\Omega_{2}) \,Y_{\lambda_{3}}(\Omega_{3})\,\times
\nonumber\\[2.5ex]
&\times\,\big\langle\,\delta\rho_{n_{1}}(\Omega_{1})\,
\delta\rho_{n_{2}}(\Omega_{2})\, \delta\rho_{n_{3}}(\Omega_{3})\,
\big\rangle \,.
\end{align}
For $\langle\,\delta\rho_{n_{1}}(\Omega_{1})\,\delta\rho_{n_{2}}
(\Omega_{2})\, \delta\rho_{n_{3}}(\Omega_{3})\,\rangle$, one can 
prove that a reasonable approximation is
\begin{align}
\label{eqD2}
\big\langle\,\delta&\rho_{n_{1}}(\Omega_{1})\,
\delta\rho_{n_{2}}(\Omega_{2})\, \delta\rho_{n_{3}}(\Omega_{3})\,
\big\rangle \,\approx \,\sum\limits_{n} \int \text{d} \Omega \,
{\rho^{(1)}(\Omega)}\,\times
\nonumber \\[1ex]
&\times\,\frac{G_{n_{1}n}(\Omega_{1},\Omega)}{\rho^{(1)}(\Omega)} \,
\frac{G_{nn_{2}}(\Omega,\Omega_{2})}{\rho^{(1)}(\Omega)} \,
\frac{G_{nn_{3}}(\Omega,\Omega_{3})}{\rho^{(1)}(\Omega)} \,,
\end{align}
where $G_{nn_{1}}(\Omega,\Omega_{1}) = G_{n_{1}n}(\Omega_{1},\Omega)$
has been used. Performing the Fourier sums of
Eq.~(\ref{eqD1}) on approximation~(\ref{eqD2}) yields
\begin{align}
\label{eqD3}
\frac{N}{(4\pi)^{3}} &\,\sum_{{\bf Q}} \,
\delta_{{\bf q}_{2}+{\bf q}_{3},{\bf q}_{1}+{\bf Q}}
\int \text{d} \Omega \,\rho^{(1)}(\Omega)\,\times
\nonumber\\[2ex]
&\times\,\frac{S({\bf q}_{1},\Omega_{1},\Omega)}{\rho^{(1)}(\Omega)} \,
\frac{S({\bf q}_{2},\Omega,\Omega_{2})}{\rho^{(1)}(\Omega)} \,
\frac{S({\bf q}_{3},\Omega,\Omega_{3})}{\rho^{(1)}(\Omega)}\,,
\end{align}
where
\begin{equation}
\label{eqD4} 
S({\bf q},\Omega,\Omega') = 4\pi \, \sum_{{\bf x}_{nn'}} 
G_{nn'}(\Omega,\Omega') \,\text{e}^{i{\bf q} \cdot {\bf x}_{nn'}} \, .
\end{equation}
Substituting
\begin{equation}
\label{eqD5}
S({\bf q},\Omega,\Omega') = \sum_{\lambda\lambda'}\hspace{0ex}'
(-i)^{l'-l} S_{\lambda\lambda'}({\bf q}) \, Y_{\lambda}(\Omega)\,
Y^{*}_{\lambda'}(\Omega')
\end{equation}
and
\begin{align}
\label{eqD6}
&\frac{S({\bf q},\Omega,\Omega')}{\rho^{(1)}(\Omega)} 
= 4\pi\int\text{d} \Omega'' \,d^{-1}(\Omega,\Omega'')\,
S({\bf q},\Omega'',\Omega') = 
\nonumber\\[1ex]
&=4\pi \sum_{\lambda}
 \sum_{\lambda'}\hspace{0ex}'
(-i)^{l'-l} \left( d^{-1}S\right)_{\lambda\lambda'}({\bf q}) \,
Y_{\lambda}(\Omega)\,Y^{*}_{\lambda'}(\Omega')\,,
\end{align}
with
\begin{subequations}
\label{eqD7}
\begin{align}
\label{eqD7a}
d(\Omega,\Omega') &= 4\pi \, \rho^{(1)}(\Omega)
\,\delta(\Omega|\Omega')\,,\\[1ex]
\label{eqD7b}
d^{-1}(\Omega,\Omega') &= \frac{1}{4\pi} \, 
\frac{\delta(\Omega|\Omega')}{\rho^{(1)}(\Omega)}
\end{align}
\end{subequations}
into Eq.~(\ref{eqD3}) and taking the $\lambda$-transforms 
as defined in Eq.~(\ref{eqD1}) afterwards we get
\begin{align}
\label{eqD8}
\big\langle \,\delta&\rho_{\lambda_{1}}^{*}({\bf q}_{1})
\,\delta \rho_{\lambda_{2}}({\bf q}_{2})\, \delta 
\rho_{\lambda_{3}}({\bf q}_{3}) \,\big\rangle \approx
\nonumber\\[2ex]
\approx&\;\frac{N}{4\pi} \,\sum_{{\bf Q}}\,
\delta_{{\bf q}_{2} + {\bf q}_{3},{\bf q}_{1} +{\bf Q}}
 \, \sum_{\lambda_{1}'}\hspace{0ex}'
\sum\limits_{\lambda_{2}'\lambda_{3}'} \,i^{l_{2}'+l_{3}'-l_{1}'} \,\times
\nonumber\\[2ex]
&\times\,c(l_{2}'l_{3}'l_{1}')\,
C(l_{2}'l_{3}'l_{1}',m_{2}'m_{3}'m_{1}')\,\times
\nonumber\\[3.5ex]
&\times\,S_{\lambda_{1}\lambda_{1}'}({\bf q}_{1}) \, (d^{-1}
S)_{\lambda_{2}'\lambda_{2}}({\bf q}_{2})\, (d^{-1}
S)_{\lambda_{3}'\lambda_{3}}({\bf q}_{3}) \, .
\end{align}
Although, the product of the last three factors of
Eq.~(\ref{eqD8}) does not look symmetric, one can show that all
three factors indeed are equivalent.
\end{document}